\documentclass[usenames,dvipsnames]{aa}
\pdfoutput=1

\ifx\pdfoutput\undefined
\usepackage{graphicx}
\else
\usepackage[pdfauthor={J.~Threlfall},citecolor=blue,linkcolor=blue, linktocpage=true,colorlinks=true]{hyperref}
\usepackage{epstopdf}
\fi 

\usepackage{natbib}
\bibpunct{(}{)}{;}{a}{}{,} 
\usepackage{txfonts}
\usepackage{subfig}	
\usepackage{epsfig}	
\usepackage{cancel}
\usepackage{multirow}
\usepackage{fixltx2e}	
\usepackage{amssymb}	
\graphicspath{{figs/}}

\newsavebox{\bigleftbox}
\newsavebox{\bigrightbox}
\DeclareRobustCommand*{\unit}[1]{\def~{\,}\ensuremath{\mathrm{\,#1}}}

\date{}			

\newcommand{\boldnabla}{\mbox{\boldmath$\nabla$}}
\def\grad{\boldnabla}
\newcommand{\vpar}{ v_{\parallel} }
\newcommand{\vperp}{ v_{\perp} }

\def\vsclsq{{{v_{scl}}^2}}

\def\tscl{{t_{scl}}}
\def\bscl{{b_{scl}}}

\def\omscl{{{\Omega_{scl}}}}

\newcommand{\new}[1]{{{#1}}}

\newcommand{\Epar}{E_{\parallel}}
\newcommand{\Eperp}{E_{\perp}}
\newcommand{\upar}{u_{\parallel}}
\newcommand{\ue}{{{\bf{u}}_E}}

\begin{document}
\title{Flare particle acceleration in the interaction of twisted coronal flux ropes}
 \titlerunning{Flare particle acceleration in the interaction of twisted coronal flux ropes}
 \author{J.~Threlfall\inst{\ref{inst1}} \and A.~W.~Hood\inst{\ref{inst1}} \and P.~K.~Browning\inst{\ref{inst2}} }
 \institute{School of Mathematics and Statistics, University of St Andrews, St Andrews, Fife, KY16 9SS, U.K. \email{\{jwt9;awh\}@st-andrews.ac.uk}\label{inst1} \and School of Physics and Astronomy, University of Manchester, Oxford Road, Manchester, M13 9PL, U.K. \email{philippa.browning@manchester.ac.uk}\label{inst2}}
 \abstract
  {}
 {The aim of this work is to investigate and characterise non-thermal particle behaviour in a three-dimensional (3D) magnetohydrodynamical (MHD) model of unstable multi-threaded flaring coronal loops.}
 {We have used a numerical scheme which solves the relativistic guiding centre approximation to study the motion of electrons and protons. The scheme uses snapshots from high resolution numerical MHD simulations of coronal loops containing two threads, where a single thread becomes unstable and (in one case) destabilises and merges with an additional thread.}  
 {The particle responses to the reconnection and fragmentation in \new{MHD simulations of} two loop threads are examined in detail. We illustrate the role played by uniform background resistivity and distinguish this from the role of anomalous resistivity \new{using orbits in an MHD simulation} where only one thread becomes unstable without destabilising further loop threads. We examine the (scalable) \new{orbit} energy gains and \new{final positions} recovered at different \new{stages of a second MHD simulation} wherein a secondary loop thread is destabilised by (and merges with) the first thread. \new{We compare these results with other theoretical particle acceleration models in the context of} observed energetic particle populations during solar flares.}
 {}
 \maketitle

\section{Introduction}\label{sec:Intro}
The physical processes which underpin particle acceleration in solar flares (across a range of scales, from the largest flares to barely observable micro- and nano-flares) present one of the most challenging unsolved problems in plasma physics to-date.
\new{The nanoflare picture, in particular, may also provide a resolution of the longstanding coronal heating problem \citep[e.g.][]{review:ParnellDeMoortel2012,review:DeMoortelBrowning2015,paper:Klimchuk2015}, with the combined effect of ubiquitous nanoflares providing the heating \citep{paper:Parker1988}.} Active region heating may be a result of `nanoflare storms' \citep{paper:Klimchuk2015} in multi-threaded coronal loops, so it is important to understand the mechanisms whereby energy release in one thread can lead to the release of energy from the neighbouring threads. 

Since non-thermal particles are a feature of \new{most flares of all scales, their presence in non-flaring active regions could provide confirmation of the nanoflare hypothesis - although} the question of how the partitioning of energy between thermal plasma and non-thermal particles (and other forms) scales with event size is not yet answered. Further modelling of the processes by which particles are accelerated will be required in order to resolve this, and to predict the expected properties of the particle distributions. The detection of energetic particles from smaller flare-like events relevant to coronal heating is beyond the capability of current hard X-ray (HXR) telescopes such as RHESSI, but the sounding rocket FOXSI provides a glimpse of the HXR emission from microflares \citep{paper:Kruckeretal2014}, and suggests that a future space-based focusing HXR instrument could provide important insights into nanoflares \citep{preprint:Christeetal2017}. Recently HXR emission signatures from a microflare have been observed using NuSTAR \citep{paper:Gleseneretal2017}\new{, which can also be used to deduce non-thermal limits for small microflares \citep{paper:Wrightetal2017}. Potential non-thermal sources have been observed for a number of years at radio frequencies \citep[see e.g.][and references therein]{review:Shibasakietal2011}, while} IRIS observations also indicate the presence on non-thermal electrons associated with nanoflare heating \citep{paper:Testaetal2014}.

Line-tied coronal loops are remarkably stable structures \citep[e.g.][]{paper:Raadu1972}, but inclusion of increasing levels of magnetic twist in coronal flux tube models (above a critical value) can allow the ideal magnetohydrodynamic (MHD) kink instability to naturally develop in previously stable loops \citep[e.g.][]{paper:HoodPriest1979}. Photospheric footpoint motions can drive this development and onset \citep[][]{paper:Gerrardetal2002}. The non-linear phase of the kink instability generates helical current sheet structures within a flux tube, leading to rapid (enhanced) reconnection, fragmentation and local heating \citep[][]{paper:Gerrardetal2001,paper:BrowningVanderLinden2003}. Subsequent build-up and dissipation of secondary current sheet structures in later stages within the same tube continue to heat the corona in a manner consistent with the nanoflare picture \citep{paper:Browningetal2008,paper:Hoodetal2009}, until the configuration approaches a minimal energy, helicity conserving Taylor state \citep{paper:Taylor1974,paper:BrowningVanderLinden2003,paper:Barefordetal2013}. This relatively simple initial concept has been further extended to incorporate additional effects, including atmospheric stratification, curvature, thermal conduction and others \citep[][]{paper:Barefordetal2016}, while producing plasma motions which qualitatively and quantitatively agree with observations \citep[][]{paper:Gordovskyyetal2016}. 

Significantly, this mechanism can also induce additional disruptions in neighbouring loop threads \citep[][]{paper:Tametal2015}, leading to a possible cascade of energy release from many threads over time \citep[][]{paper:Hoodetal2016} in a manner also consistent with helicity conserving Taylor relaxation \citep[][]{paper:Hussainetal2017}. This is important, because it means energy can be released from twisted threads even if they are well below the kink instability threshold. Instability is needed in only one thread, which then triggers an avalanche of heating (and, as we show, particle acceleration) in its neighbours.

The characteristic behaviour of particle orbits in the vicinity of isolated current sheets during reconnection events have also been studied for some years \citep[e.g.][]{paper:Turkmanietal2005,paper:Onofrietal2006,paper:Gordovskyyetal2010a,paper:Gordovskyyetal2010b, paper:GordovskyyBrowning2011, paper:GordovskyyBrowning2012}. Models of coronal loops which destabilise in the manner described above provide a convenient global framework to compare orbit results with observations, while also allowing extensions which account for the impact of atmospheric stratification, collisions and instrumentational effects in the model
\citep[e.g.][]{paper:Gordovskyyetal2013,paper:Gordovskyyetal2014, paper:Pintoetal2016}. \new{Underpinning this approach is the idea that magnetic reconnection is generically associated with parallel electric fields \citep[e.g.][]{paper:Schindleretal1988,paper:HesseSchindler1988,paper:Schindleretal1991}, which are strong candidates to accelerate particles \citep{paper:Janvieretal2015}.}

Of particular relevance for this investigation is the work of \citet{paper:GordovskyyBrowning2011,paper:GordovskyyBrowning2012}, who calculate electron and proton orbits in a single twisted flux tube within which the ideal kink instability develops, and study the particle acceleration efficiency for a number of magnetic resistivity profiles. With the relatively recent `discovery' of multiple thread eruptions triggered by a single kink unstable thread \citep[termed an `avalanche' model, e.g. by][]{paper:Hoodetal2016}, it seems pertinent to ascertain how the energetic particle properties (including energies and impact sites) may change when the eruption takes place in several (neighbouring) threads over time.

The primary objective of this investigation is to determine particle orbit behaviour during an event  containing multiple magnetic threads. Unlike previous studies, an instability in one of the threads can lead to the destabilisation of others. We have focussed on a system containing one unstable thread and one stable thread; however, similar behaviour would be expected in multi-thread configurations \citep[such as][for example]{paper:Hoodetal2016}. Our objective is to analyse how the particle orbit response differs when a series of threads destabilise and merge, compared to the single destabilisation of one thread alone, and what factors affect this response. 

Our paper is organised as follows:
In Section~\ref{sec:model} we discuss the model itself, which is divided into two parts. In Section~\ref{subsec:MHDmodel}, we describe the global MHD field model (containing two neighbouring flux tubes, one of which is driven beyond marginal stability). In Section~\ref{subsec:GCmodel}, we describe our particle orbit model, which takes input from the MHD model. 
The preliminary stage of the investigation, Section~\ref{sec:oneunstable}, revisits the case where only a single flux tube destabilises and fragments, where our focus is to investigate the effect of background and anomalous resistivity on the particle orbits. Then we consider the orbit response to a case where a second loop instability is triggered by the first, in Section~\ref{sec:twounstable}. We discuss our results in Section~\ref{sec:disc}, before presenting conclusions and possible areas of future study in Section~\ref{sec:conc}.

\section{Model setup}\label{sec:model}

\subsection{MHD model}\label{subsec:MHDmodel}
MHD simulations of a kink unstable loop interacting and disrupting a neighbouring ideally stable loop have been performed by \citet{paper:Tametal2015} and \citet{paper:Hoodetal2016}. A discussion of the equations
solved and the parameters used are given in \citet{paper:Tametal2015}, while the Lagrangian remap code, {\textit{Lare3D}}, used here is described in \citet{paper:LareXd2001}. The non-linear evolution of the
line-tied kink instability has been studied by several authors \citep[e.g.][]{paper:Batyetal1998,paper:Lionelloetal1998,paper:Gerrardetal2001,paper:Gerrardetal2002, paper:Browningetal2008, paper:Hoodetal2009}. During the non-linear ideal phase of the instability, a helical current sheet forms in the unstable magnetic loop. Once reconnection starts, this current sheet fragments, forming many small current sheets within the unstable loop. The formation of these small current sheets helps the magnetic field to relax towards a lower energy state. One consequence of the relaxation process is that the cross section of the unstable loop expands. This expansion allows this loop to interact with a neighbouring stable loop. \citet{paper:Tametal2015} showed that if the stable loop is located sufficiently near to the unstable loop, the stable loop can be disrupted. On the other hand, if the stable loop is sufficiently far from the unstable one, there is no interaction. When a nearby stable loop is disrupted, its stored magnetic energy can be released. The interaction creates a current sheet in the stable loop and, once reconnection starts, the stable loop is attracted into the unstable loop. This dynamical merging results in the formation of more current sheets and more reconnection. The two loops merge, relaxing to a single, weakly twisted loop. \citet{paper:Hoodetal2016} showed that the interaction of loops can continue, with yet more stable loops being disrupted, resulting in an MHD avalanche.

The MHD simulations produce the 3D temporal evolution of all the plasma and magnetic field variables. The {\textit{Lare3D}} code uses dimensionless variables, which we base on a magnetic field strength, $B_0 = 10$G, a length, $L = 10^{6}$m, and a mass density, $\rho_0 = 3.3 \times 10^{-12}\unit{kgm}^{-3}$. Thus, the typical Alfv\'en speed is $V_A = 491$km s$^{-1}$ and the typical time is $t=2$s. The reference current density is $j_0 = B_0/(\mu L) = 8\times 10^{-4}\unit{Am}^{-2}$ and the magnetic diffusivity is $\eta_0 = 5 \times 10^{11}\unit{m}^{2}\unit{s}^{-1}$. Time units are quoted in Alfv\'en times, $\tau_A$. The domain considered is as follows: $-2 \le \bar{x} \le 4$,$-2 \le \bar{y} \le 2$ and $-10 \le \bar{z} \le10$ (where barred quantities represent non- dimensional variables in the numerical domain). Both loops have a radius of unity.

A key question that is not addressed by MHD simulations is how the instability, reconnection and any subsequent disruption influences particle acceleration. Particle acceleration is driven by the electric field and the component parallel to the magnetic field depends strongly on the form of the resistivity chosen. While a small background resistivity is useful for energy conservation, it often results in (artificially) large particle acceleration. An anomalous resistivity, triggered when the current rises above a critical value, allows the formation of strong currents and ensures that the reconnection remains spatially local. In dimensionless values, the two resistivities are $\bar{\eta}_{bkg} = 5 \times 10^{-5}$, for the uniform background value and $\bar{\eta} = 0.001$ when $|j| \ge 5$ for the anomalous value. 
Following the work of \citet{paper:Tametal2015}, we consider two simulations: the first simulation has only a single unstable loop (Section~\ref{sec:oneunstable}); and the second simulation involves the disruption of a stable loop by the unstable one (Section~\ref{sec:twounstable}). The particle orbit calculations for the single loop were repeated, once with the background resistivity switched off ($\bar{\eta}_{bkg}= 0$) and once with it on. The simulations involving the disruption of the stable loop were performed with no background resistivity.

\subsection{Relativistic particle dynamics}\label{subsec:GCmodel}
We investigate energetic electron and proton properties using the test-particle approach. This is valid if the population of non-thermal energetic particles is small compared with the thermal background population which generates the MHD fields. Having established the details of the MHD simulations (whose snapshots we will use as a general environment within which we will initialise particle orbits) we now outline how each orbit will be calculated.
We use the guiding centre approximation \citep[or GCA, derived in][]{book:Northrop1963} to study the evolution of test-particle orbits, using an identical approach to that used in various recent studies of orbit behaviour in a range of environments \citep{paper:Threlfalletal2016b,paper:Threlfalletal2016a,paper:Borissovetal2016,paper:Threlfalletal2017a}, including studies of separator reconnection, non-topological reconnection, and indeed full MHD simulation snapshots of an entire active region. This approach is not unique \citep[e.g.][]{paper:GordovskyyBrowning2011,paper:GordovskyyBrowning2012,paper:Gordovskyyetal2014}. In seeking to extend the work of \citet{paper:GordovskyyBrowning2011,paper:GordovskyyBrowning2012}, we use a similar approach, in order that we may compare like-for-like results. Details of our implementation of this method are readily available, having been fully outlined in previous investigations \citep[e.g.][]{paper:Threlfalletal2017a}. For brevity, we will only briefly review key details.

We calculate the movement of the guiding centre ${\bf{R}}$ of a particle with rest-mass $m_0$, charge $q$ and relativistic magnetic moment, $\mu_r$. The orbit is controlled by a magnetic field ${\bf{B}}$ (with magnitude $B=|{\bf{B}}|$ and unit vector ${\bf{b}}={\bf{B}}/B$) and an electric field ${\bf{E}}$. It responds according to the relativistic GCA equations:
\begin{subequations}
 \begin{align}
  \frac{d{u_\parallel}}{dt}&=\frac{d}{dt}\left(\gamma\vpar\right)=\gamma\ue\cdot{\frac{d{\bf{b}}}{dt}}+\omscl\tscl E_\parallel-\frac{\mu_r}{\gamma}\frac{\partial{B}}{\partial s}, \label{eq:Rnorm1} \\
  {\bf\dot{R}_\perp}&=\ue+\frac{\bf{b}}{B^{\star\star}}\times\left\lbrace \frac{{1}}{\omscl\tscl}\left[ \frac{\mu_r}{\gamma}\left( \grad{B^\star}+ \frac{\vsclsq}{{c^2}}\ue\frac{\partial B^\star}{\partial t}\right)\right.\right. \nonumber \\ 
   &\qquad\quad\qquad\qquad\left.\left. +u_\parallel\frac{d{\bf{b}}}{dt}+\gamma\frac{d\ue}{dt}\right]+\frac{\vsclsq}{{c^2}}\frac{u_\parallel}{\gamma}{E_\parallel}\ue \right\rbrace, \label{eq:Rnorm2} \\ 
  \frac{d\gamma}{dt}&=\frac{\vsclsq}{{c^2}}\left[\omscl\tscl\left({\bf\dot{R}_\perp}+\frac{u_\parallel}{\gamma}{\bf{b}}\right)\cdot{\bf{E}}+\frac{\mu_r}{\gamma}\frac{\partial B^\star}{\partial t}\right],   \label{eq:Rnorm3} \\
  \mu_r&=\frac{\gamma^2{\vperp^2}}{B}, \label{eq:Rnorm4}  
 \end{align}
 \label{eq:rel_norm} 
\end{subequations}
where $B^{\star}$ and $B^{\star\star}$ modify the original field strength $B$ through
\[
 B^\star=B\left( 1-\frac{1}{c^2}\frac{{\Eperp}^2}{B^2}\right)^{\frac{1}{2}} , \qquad B^{\star\star}=B\left(1-\frac{1}{c^2}\frac{{\Eperp}^2}{B^2}\right).
\]

Particle drifts (largest of which is the ${E}\times{B}$ drift, which causes the orbit to drift at a velocity $\ue={\bf{E}}^{\star}\times{\bf{b}}/B$) affect the guiding centre motion. Velocity and electric field components aligned with the magnetic field (henceforth known as parallel velocity and parallel electric field) are $\vpar(={\bf{b}}\cdot{\dot{\bf{R}}})$ and $\Epar(={\bf{b}}\cdot{\bf{E}})$, respectively. $\vperp$ is the gyro-velocity, $\dot{\bf{R}}_\perp(=\dot{\bf{R}}-\vpar{\bf{b}})$ is the perpendicular component of guiding centre velocity, and $s$ is an arc-length parallel to the magnetic field. We also, for simplicity, define a relativistic parallel velocity $\upar(=\gamma\vpar)$, for the usual Lorentz factor $\gamma \left(= c/\left(c^2-v^2\right)^{1/2}\right)$.

In this work, we examine the motion of electrons and protons. Hence, the rest mass and charge of our test-particles are either fixed to be $m_0=m_e=9.1\times10^{-31}$\unit{kg} and $q=e=-1.6022\times10^{-19}$\unit{C} for electrons or  $m_0=m_p=1.67\times10^{-27}\unit{kg}$ and $q=|e|=1.6022\times10^{-19}\unit{C}$ for protons.

Equations~(\ref{eq:rel_norm}) have been non-dimensionalised. Dimensional values are obtained using appropriate values, which were chosen to be relevant in a solar coronal context. Times are normalised by a single Alfv\'en time from the MHD simulations ($\tscl=\tau_A=2$\unit{s}). The remaining normalising parameters for the GCA scheme use the same normalisation as the MHD simulations, yielding a dimensionless electron or proton gyro-frequency $\omega(={q\,\bscl}{m_0}^{-1})$. The factor $\omega\tscl$ should be large in order for guiding centre theory to be valid (as drift terms appear in Eqs.~\ref{eq:rel_norm} of the order of the inverse of this parameter). For the parameters mentioned, $\omega\tscl=-3.5\times10^{8}\unit{rads}$ for electrons and $1.92\times10^{5}\unit{rads}$ for protons, justifying our use of the guiding centre approach here. 
\begin{figure*}[t]
 \centering
 \subfloat[$t=0\tau_{A}$]{\label{subfig:1Lt0B}\resizebox{0.32\textwidth}{!}{\includegraphics[clip=true, trim=110 100 110 190]{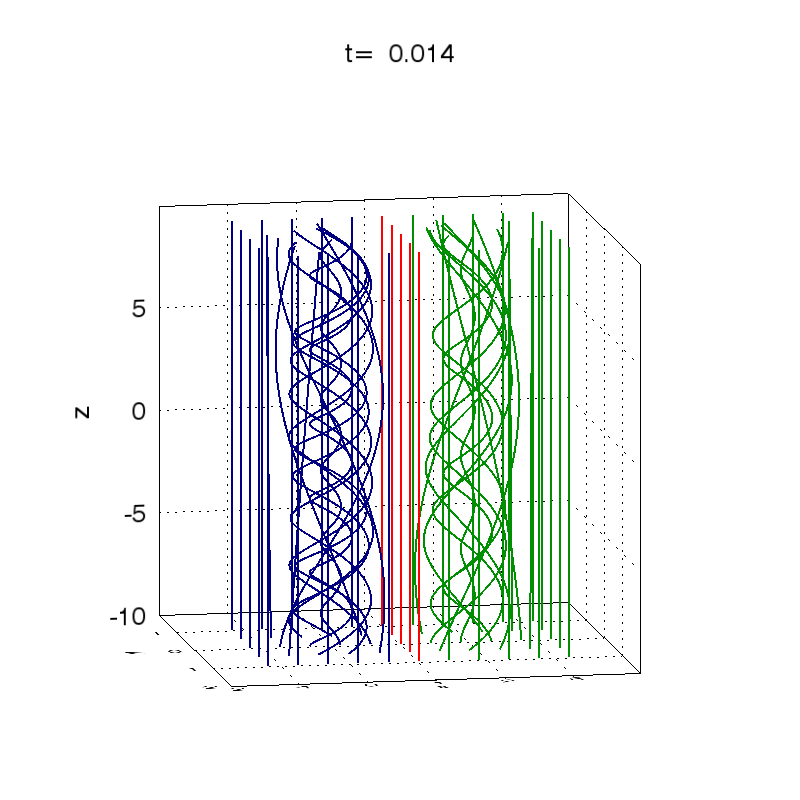}}}
 \subfloat[$t=65\tau_{A}$]{\label{subfig:1Lt1B}\resizebox{0.32\textwidth}{!}{\includegraphics[clip=true, trim=110 100 110 190]{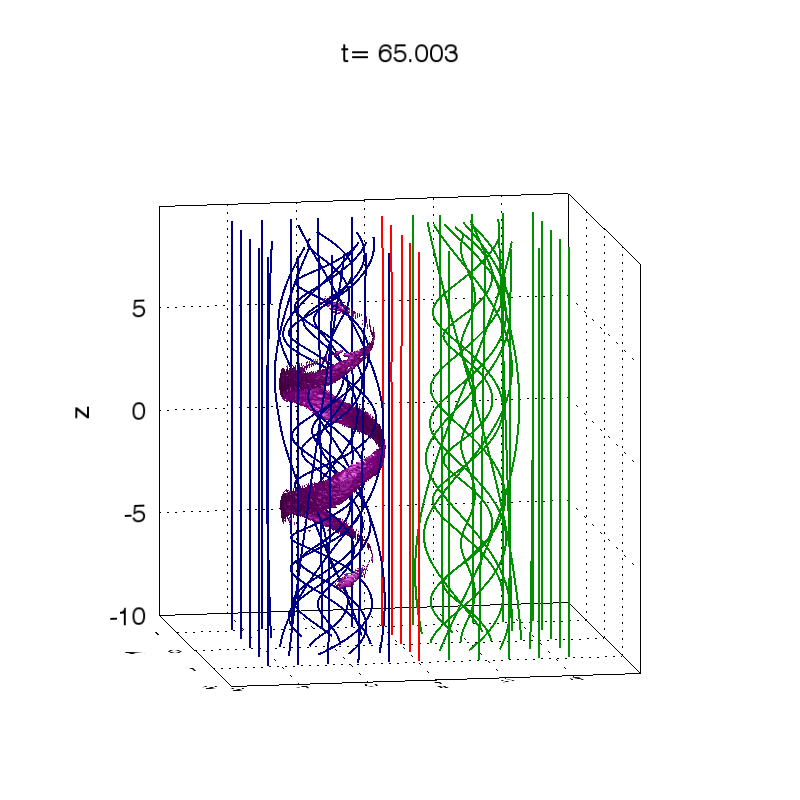}}}
 \subfloat[$t=180\tau_{A}$]{\label{subfig:1Lt2B}\resizebox{0.32\textwidth}{!}{\includegraphics[clip=true, trim=110 100 110 190]{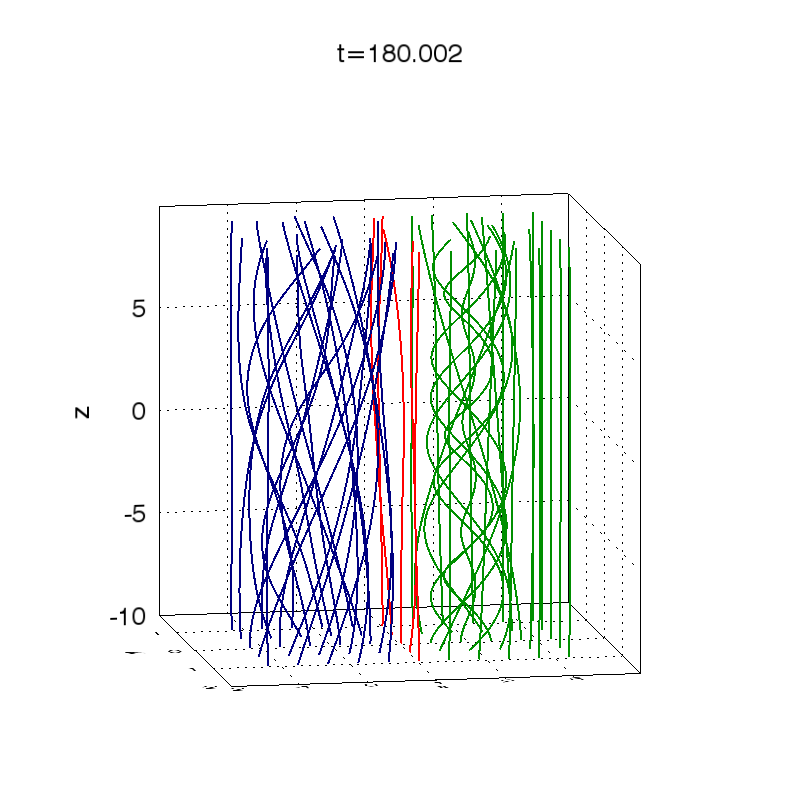}}}
  \caption{Single loop thread destabilises: Images show specific interpolated magnetic field lines (traced from the base of the simulation domain, at $\bar{z}=-10$ in non-dimensional units), colour-coded by location in $\bar{x}$, at different stages of the experiment. \new{Blue} field lines are traced from regions where $\bar{x}<-0.1$, \new{red} field lines are traced from $\bar{x}=0$ only, while \new{green} field lines are traced from $\bar{x}>0.1$. Hence in \protect\subref{subfig:1Lt0B} (prior to any reconnection) blue field lines are initially associated with the left-hand flux tube and green field lines with the right-hand tube, separated by red field lines. Purple isosurfaces (where present) indicate regions of current above the critical value. Each image is associated with the time shown in the caption.}
 \label{fig:oneloop_why}
\end{figure*}

In order to determine the particle behaviour, Eqs.~(\ref{eq:rel_norm}) are evolved in time using a fourth order Runge-Kutta scheme with a variable timestep and compared with solutions derived from a fifth order Runge-Kutta scheme to ensure accuracy. The local electric and magnetic environment for each orbit is determined using several snapshots of our numerical MHD experiments, where the local (sub-grid) field values are found through linear interpolation to the orbit location, in both space and time. 
\begin{figure}[t]
 \centering
  \resizebox{0.49\textwidth}{!}{\includegraphics[clip=true, trim=30 10 0 0]{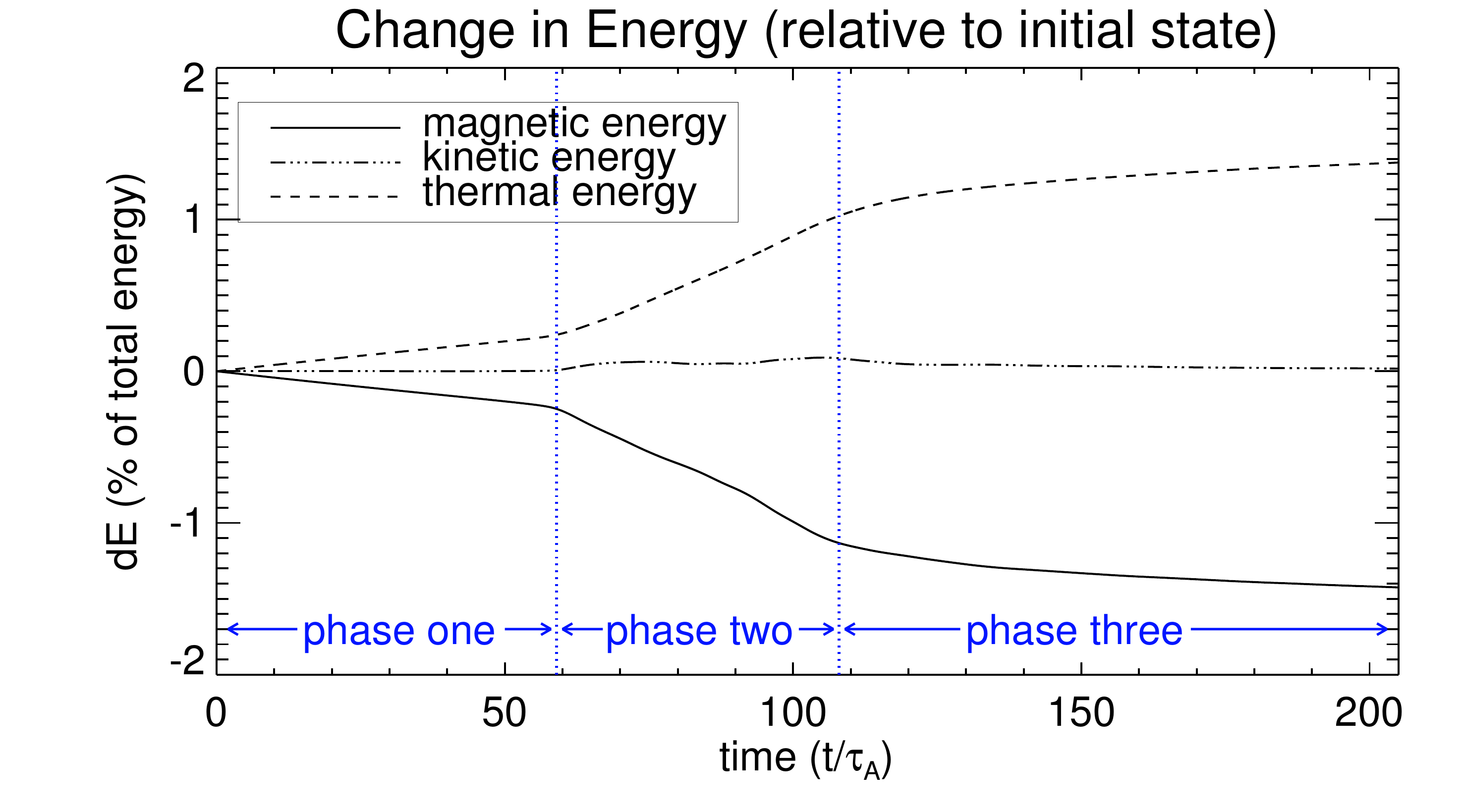}}
  \caption{Single loop thread eruption: Energy partition over time (in MHD simulation including background and anomalous resistive effects). \new{The energetics can be divided into three distinct phases (outlined in blue), while the key to individual components is given in the legend}.}
 \label{fig:dEn1}
\end{figure}

The high spatial resolution in our MHD grid prevents us from using more than five MHD snapshots for each set of orbit calculations (as each high resolution snapshot is held in memory for the duration of one set of orbit calculations). The snapshots are themselves separated in time by several temporal orders of magnitude, compared to the timescale of a single gyroperiod. In order to study different stages of each MHD experiment, we will load in five consecutive snapshots from different times in the MHD simulation per set of orbit calculations. Each orbit will be initialised at the time of the first snapshot and will continue until either a domain boundary is reached by the orbit or the orbit lifetime exceeds the time associated with the fifth snapshot in the sequence. Thus the maximum orbit lifetime in this environment is approximately $25\tau_A$ ($50\unit{s}$). In this way, orbits which remain in the domain will experience long-term changes in the background MHD environment and react accordingly, without the significant computational resources required to load in all MHD snapshots and simulate individual orbits for unfeasibly large numbers of gyroperiods.

\section{Particle acceleration in a single thread: effects of resistivity profile}\label{sec:oneunstable}
We begin by considering the case where only a single thread becomes unstable (i.e. without disrupting other loop threads). This simulation uses the first of two configurations outlined in Section~\ref{subsec:MHDmodel}, and utilises both (uniform) background resistive effects together with a (non-uniform) anomalous resistivity, triggered wherever the current is locally greater than a critical value.

Figure~\ref{fig:oneloop_why} illustrates how this system evolves in the case with non-zero background resistivity ($\eta_{\rm{bkg}}$). The left-hand flux tube (seen using \new{blue} field lines in e.g. Fig.~\ref{subfig:1Lt0B}) is perturbed and becomes kink unstable. This ultimately leads to the formation of a large helical current sheet (seen in purple in Fig.~\ref{subfig:1Lt1B}) in this tube. Anomalous resistivity acts upon this current sheet, leading to enhanced heating and a restructuring of the left-hand tube. In the later stages of the experiment, little remains of the left-hand tube, while the right-hand tube remains unaffected (shown by \new{green} field lines in e.g. Fig.~\ref{subfig:1Lt2B}).

Figure~\ref{fig:dEn1} illustrates how the energy components of this system change with time (relative to their initial state). These changes define and distinguish between three distinct `phases' of the experiment. We will compare the orbit behaviour in each phase to uncover how the particle response changes during the MHD experiment. Phase~1 is marked by a continuous slow decrease in magnetic energy from its initial value, primarily balanced by an increase in the amount of internal energy (or local heating) in the system. This phase occurs from $t=0\sim60\tau_A$ (according to Fig.~\ref{fig:dEn1}) and is dominated by $\eta_{\rm{bkg}}\neq0$ effects. The background resistivity steadily releases energy from the magnetic field to (Ohmically) heat the surrounding plasma. No current is recorded which surpasses the critical value ($j_{\rm{crit}}$) in Phase~1, guaranteeing that anomalous resistive effects are not responsible for the energy change during this phase. A helical current sheet finally surpasses $j_{\rm{crit}}$ between $t=55\tau_A$ and $60\tau_A$ (seen in Fig.~\ref{subfig:1Lt1B}). This current sheet reconnects and fragments, followed by several secondary current sheets which also form (above $j_{\rm{crit}}$), reconnecting and fragmenting for a further $\sim50\tau_A$. During this second phase, earlier current sheets are typically associated with the edge of the unstable flux tube, while later in Phase~2 current sheets form throughout the unstable flux tube region. As a result, Phase~2 maintains an enhanced rate of energy loss compared to Phase~1. The increase can be attributed to the activation of a larger anomalous resistivity (acting on laminar current sheets above the critical value) in combination with the (lower, uniform) background resistivity (acting on currents throughout the domain). Phase~2 ends after $t\approx105\tau_A$. In Phase~3, the magnetic energy loss rate returns to a gradient similar to that observed in Phase~1. In this third and final phase peak values of current return below $j_{\rm{crit}}$ (as evidenced by Fig.~\ref{subfig:1Lt2B}, where no purple isosurfaces are visible). Phase 3 sees the system return to one primarily heated by background resistive effects, particularly in the final stages.
\begin{figure*}[t]
 \centering
  \subfloat[Final electron positions]{\label{subfig:1Lt0pos}\resizebox{0.21\textwidth}{!}{\includegraphics[clip=true, trim=25 20 30 10]{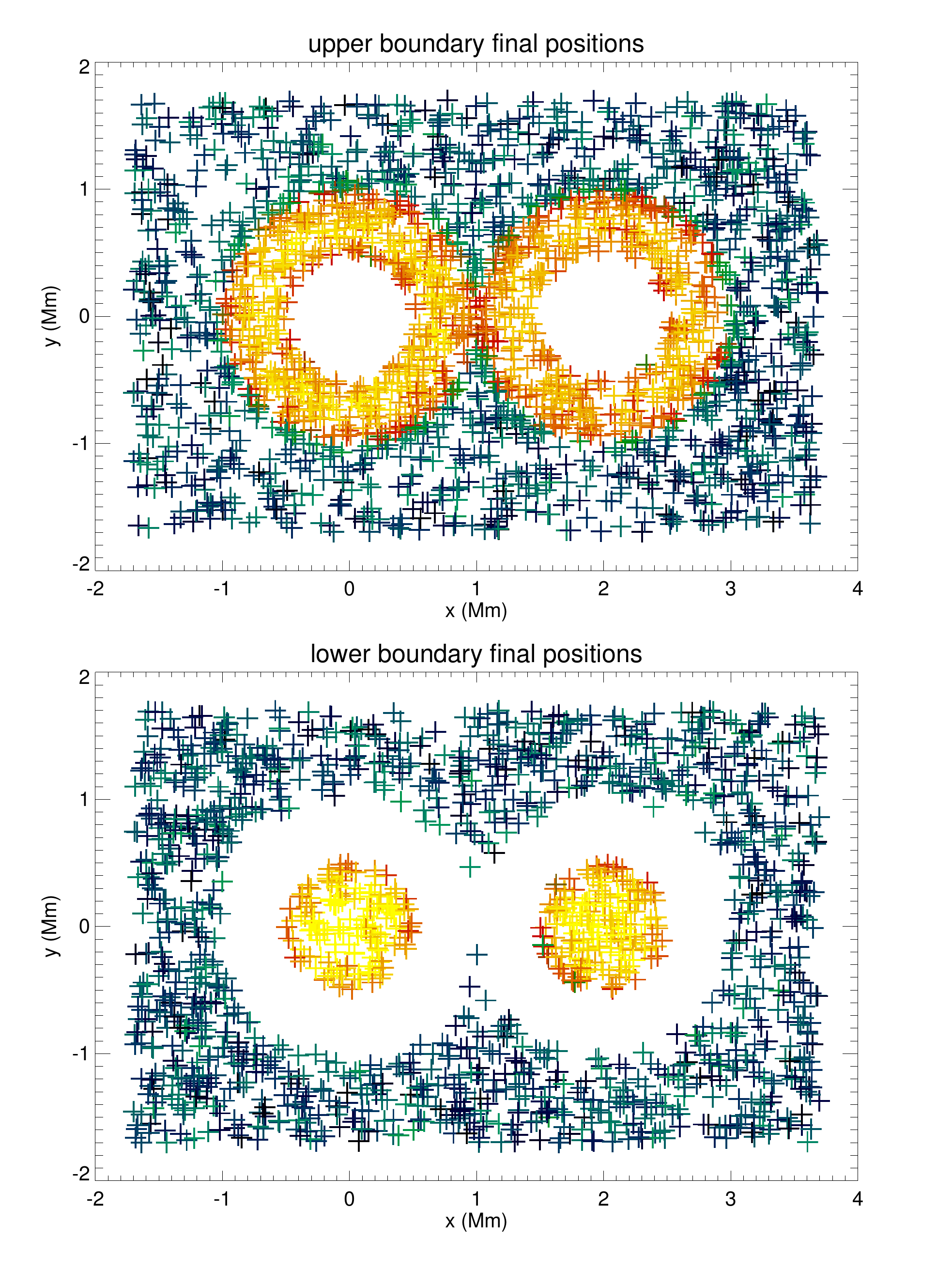}}}
 \subfloat[Final electron positions ($\eta_{\rm{bkg}}=0$)]{\label{subfig:1Lt0pos2}\resizebox{0.21\textwidth}{!}{\includegraphics[clip=true, trim=25 20 30 10]{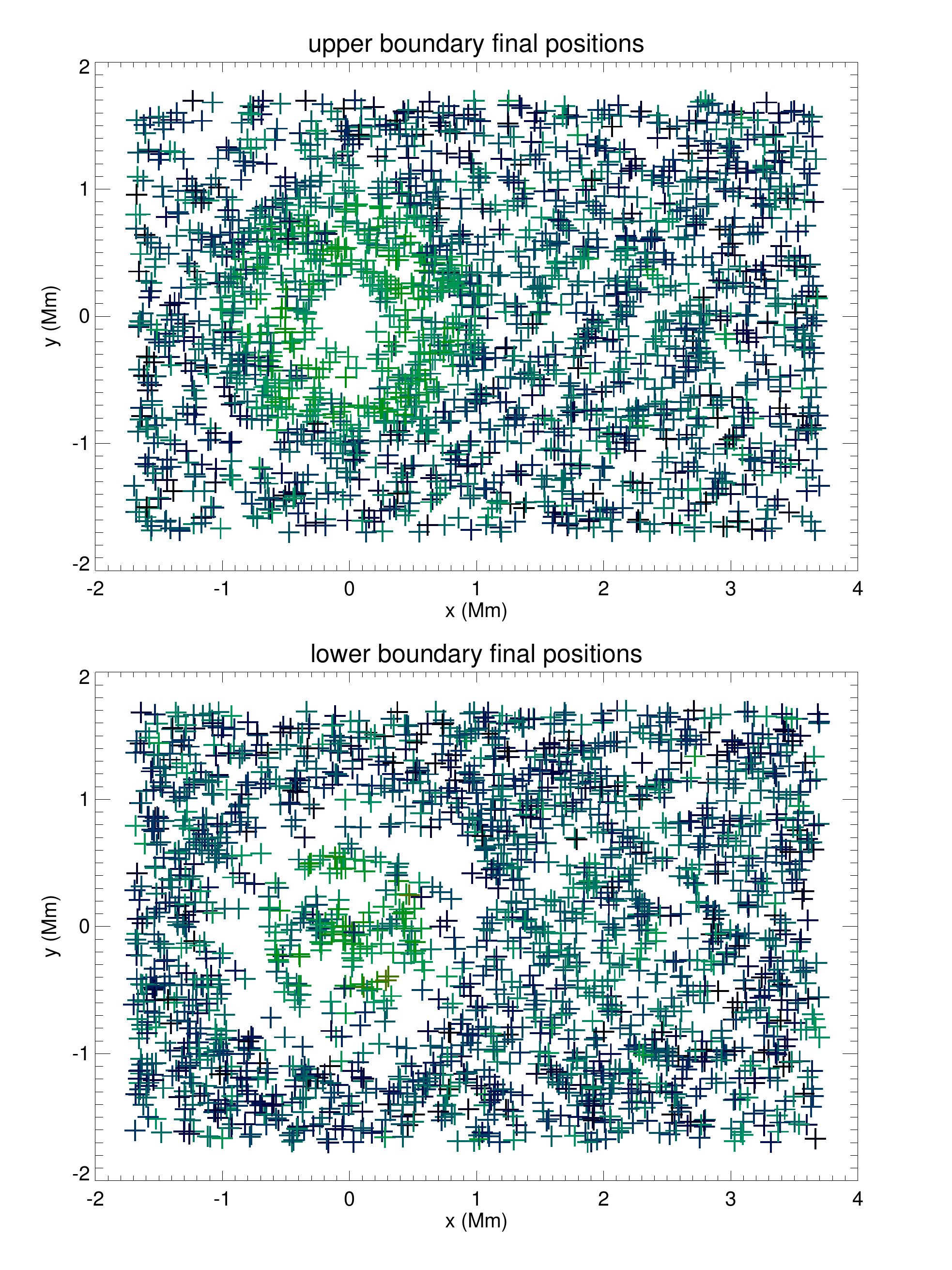}}}
 \subfloat[Final proton positions]{\label{subfig:1Lt0posP}\resizebox{0.21\textwidth}{!}{\includegraphics[clip=true, trim=25 20 30 10]{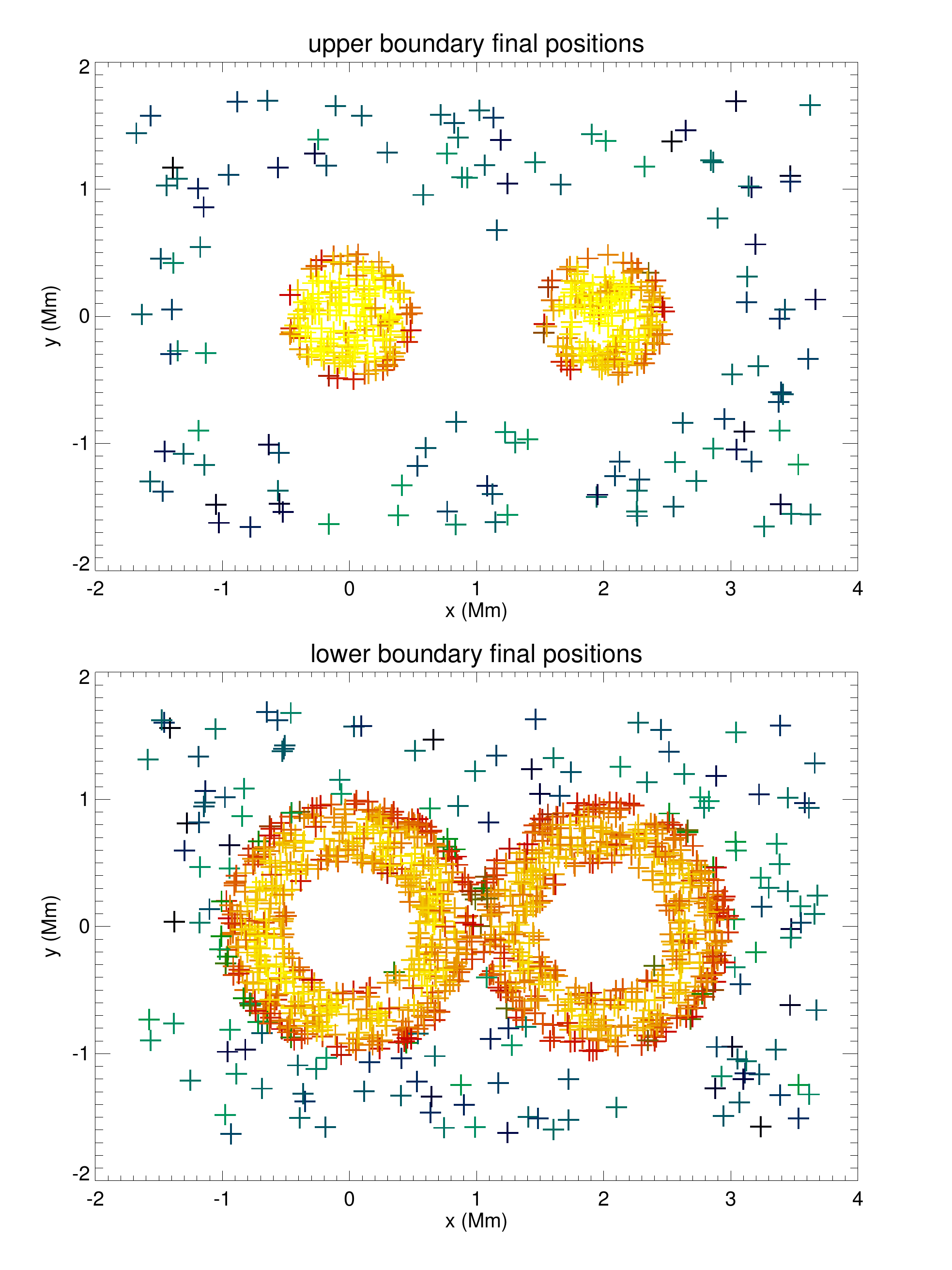}}}
 \subfloat[Final proton positions ($\eta_{\rm{bkg}}=0$)]{\label{subfig:1Lt0posP2}\resizebox{0.21\textwidth}{!}{\includegraphics[clip=true, trim=25 20 30 10]{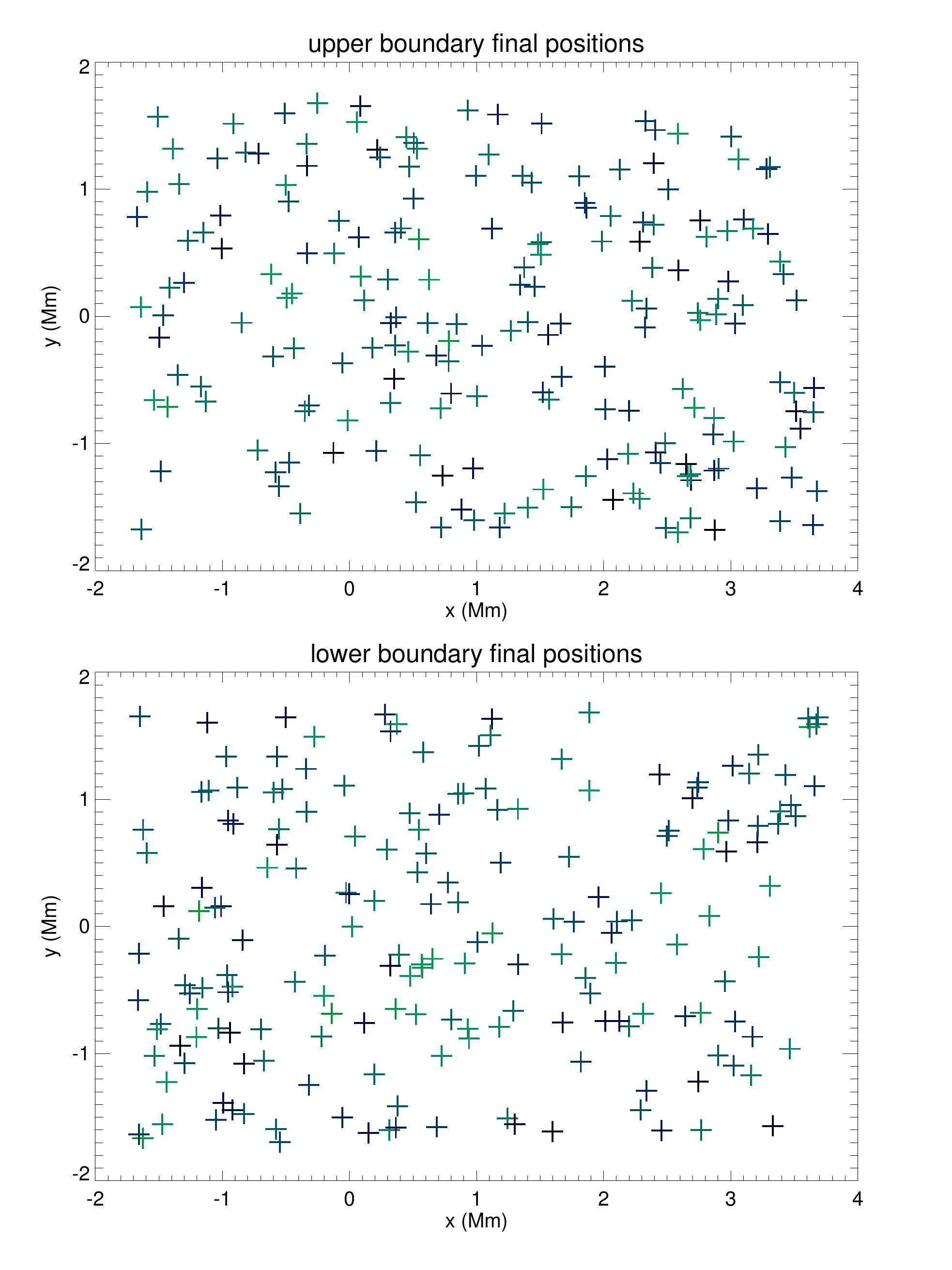}}}
  \resizebox{0.11\textwidth}{!}{\includegraphics{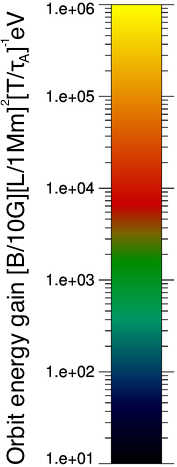}}
  \caption{Single thread destabilises, Phase~1. Impact sites (and associated energies) at upper and lower boundaries for case where electron orbits \protect\subref{subfig:1Lt0pos} include or \protect\subref{subfig:1Lt0pos2} omit background resistive effects. \protect\subref{subfig:1Lt0posP} and \protect\subref{subfig:1Lt0posP2} show equivalent proton impact sites in both cases. For key to impact site energy, see colour bar.}
 \label{fig:oneloopt0}
\end{figure*}

We will now examine the particle response to each of these phases in detail. From Fig.~\ref{fig:dEn1}, it is clear that both anomalous and background resistivity play different roles in the various phases. What is unclear is how this would affect particle orbit behaviour during each phase \citep[though the role of different resistivity models was studied in some detail for electrons and protons in a single unstable flux tube model by][]{paper:GordovskyyBrowning2012}. To distinguish the effective role of each resistivity upon individual particle behaviour in our work, we perform a set of orbit calculations in each phase, but repeat each set of calculations for two distinct cases, where the orbit calculations include or omit $\eta_{\rm{bkg}}$. Thus, during each phase of the MHD evolution, we will be able to disentangle the effects of both components of resistivity in these simulations.

Each set of orbits considered here contains $4096$ particles, having random initial positions within our simulation domain and random initial pitch angles (from $0$ to $\pi$). The initial kinetic energy of each orbit is determined by a Maxwellian distribution, peaking at a temperature close to $1$MK. \new{In principle, each orbit should be seeded with the thermal energy corresponding to the temperature of the MHD simulation region into which they are inserted. In practice, previous works \citep[e.g.][]{paper:Threlfalletal2017a} have explored a range of initial conditions and found that the initial energy is relatively unimportant, with the final state often dominated by the strength and extent of the reconnection electric fields (if present).}

\subsection{Phase 1: initial acceleration}\label{subsec:p1}
\begin{figure}[t]
 \centering
 \resizebox{0.48\textwidth}{!}{\includegraphics[clip=true, trim=10 5 10 10]{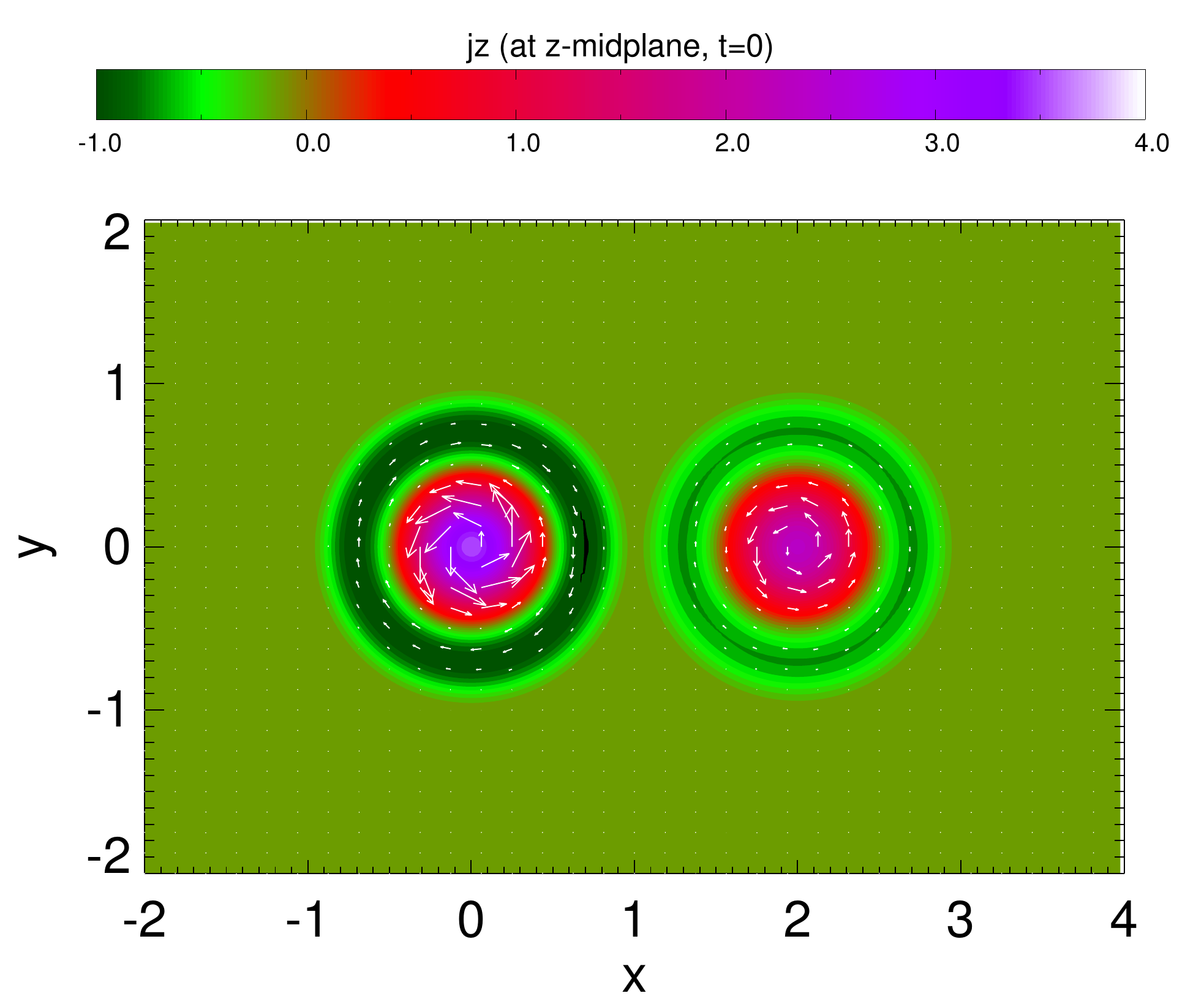}}
  \caption{Single thread destabilises. Initial current density structure in the $(x,y,z=0)$ plane. Coloured contours indicate vertical current ($j_z$), with key given by colour bar. Overlaid arrows indicate orientation and strength of planar current, $(j_x,j_y)$. }
 \label{fig:jt0}
\end{figure}

\new{Background resistivity, $\eta_{\rm{bkg}}$, dominates the particle orbit response in Phase~1. A comparison of the final positions of orbits passing through the top or bottom boundaries in Fig.~\ref{fig:oneloopt0} reveals that the background resistivity (which is small compared to the anomalous resistivity) is sufficient to accelerate both protons and electrons to high energies. Acceleration signatures can clearly be seen in the cases where background resistive effects were included, manifesting in the beams of red/orange final positions in Fig.~\ref{subfig:1Lt0pos} for electrons and Fig.~\ref{subfig:1Lt0posP} for protons. These these signatures appear identically on opposite boundaries when comparing both species, comprising of narrow beams at the tube centre surrounded by a halo of similar energy final positions on the opposite boundary. The structure of local current density, $\bf{j}$, (sampled in a midplane cut in Fig.~\ref{fig:jt0} for example) is clearly responsible for these structures. This current commonly forms two regions of opposite orientation per flux tube. The magnetic field remains uni-directional in both flux tubes, and hence the local parallel electric field reverses sign when moving from the interior of each tube to the exterior, resulting in the beam and halo structures seen in Figs.~\ref{subfig:1Lt0pos} and~\ref{subfig:1Lt0posP}.}

\begin{figure}[t]
 \centering
 \resizebox{0.49\textwidth}{!}{\includegraphics[clip=true, trim=60 15 20 37]{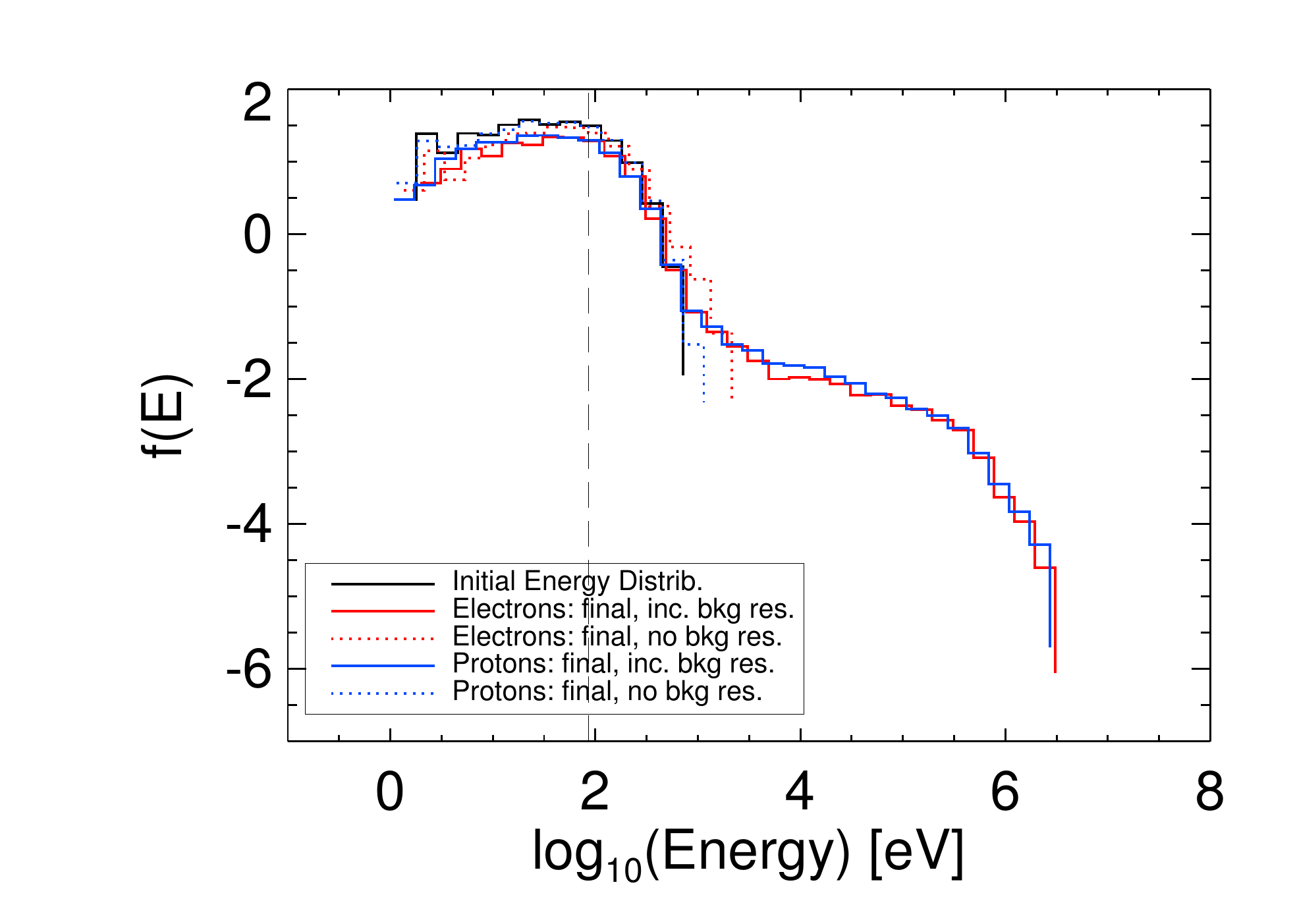}}
  \caption{Single thread destabilises, Phase~1. \new{Final energy distributions for both electron (red) and proton (blue) orbit calculations during this phase, with the calculations either including (solid lines) or omitting (dotted lines) the effects of background resistivity. An example of the initial energy distribution used by either species can be seen as the black solid histogram.}}
 \label{fig:1Lt0En}
\end{figure}

\new{The maximum electron energy in Fig.~\ref{subfig:1Lt0pos} is $3.1\unit{MeV}$, while the equivalent proton value in Fig.~\ref{subfig:1Lt0posP} is $3.07\unit{MeV}$. The pattern of energy gains is closely matched in both flux tubes. Final positions outside the core and halo structures are determined by the initial (random) distributions of position, pitch angle and (Maxwellian) energy. 
In contrast, lacking the core and halo structures, the maximum energies achieved in Figs.~\ref{subfig:1Lt0pos2} and~\ref{subfig:1Lt0posP2} are much smaller ($2.7\unit{keV}$ and $1.4\unit{keV}$ respectively). Orbits which neglect $\eta_{\rm{bkg}}$ remain at approximately Maxwellian energies. Overall, far fewer unaccelerated protons reach either top and bottom boundaries before five MHD snapshots have been used, compared to electrons, due to the difference in mass.

Electron and proton energy distributions in Fig.~\ref{fig:1Lt0En} reflect these findings; high energy tails are present whenever background resistive effects are included (solid red/blue histograms). When omitted, both electron and proton final spectra (dashed histograms) are much closer matches to the initial (Maxwellian) distribution.}

It should be noted here that the particle spectra seen in Fig.~\ref{fig:1Lt0En} (and those for subsequent phases or experiments) are calculated instantaneously by counting the number of particles in (and weighted by the width of) specific kinetic energy bins, as a general guide to illustrate how energised each orbit population has become. This is quite different from observational particle spectra, which are based on particle fluxes through a certain area in a given time and at a certain energy range. Our populations only cover a very small fraction of the range of possible initial conditions; therefore, we urge caution when comparing aspects of our recovered spectra with observationally derived spectra.

\subsection{Phase 2: critical current influence}\label{subsec:p2}
\new{Large laminar current sheet(s) form above the critical value in Phase~2. For brevity, we will henceforth only visually illustrate the final orbit positions of protons whose orbits omit background resistivity: electron and proton results are closely matched, while orbits that include $\eta_{\rm{bkg}}$ often behave as seen in Phase~1 in Fig.~\ref{subfig:1Lt0pos}.}
\begin{figure*}[t]
 \centering
   \begin{minipage}[b]{0.48\textwidth}
    \subfloat[Proton final positions, energies \& field structure ($\eta_{\rm bkg}=0$)]{\label{subfig:1Lt1pos}
    \resizebox{\textwidth}{!}{\includegraphics{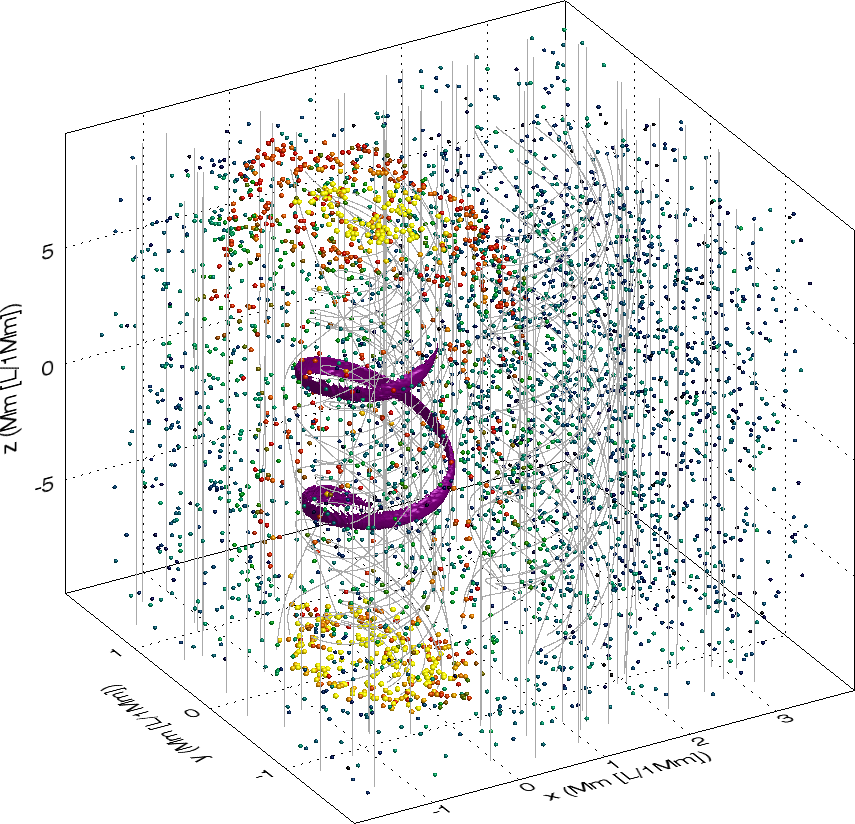}}}
   \end{minipage}
  \begin{minipage}[b]{0.45\textwidth}
  \resizebox{\textwidth}{!}{\includegraphics{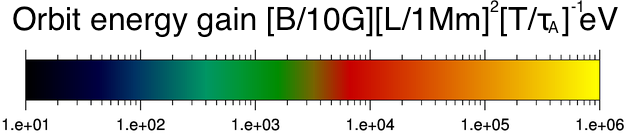}}\\
  \subfloat[Energy distributions]{\label{subfig:1Lt1sp}\resizebox{\textwidth}{!}{\includegraphics[clip=true, trim=55 15 30 20]{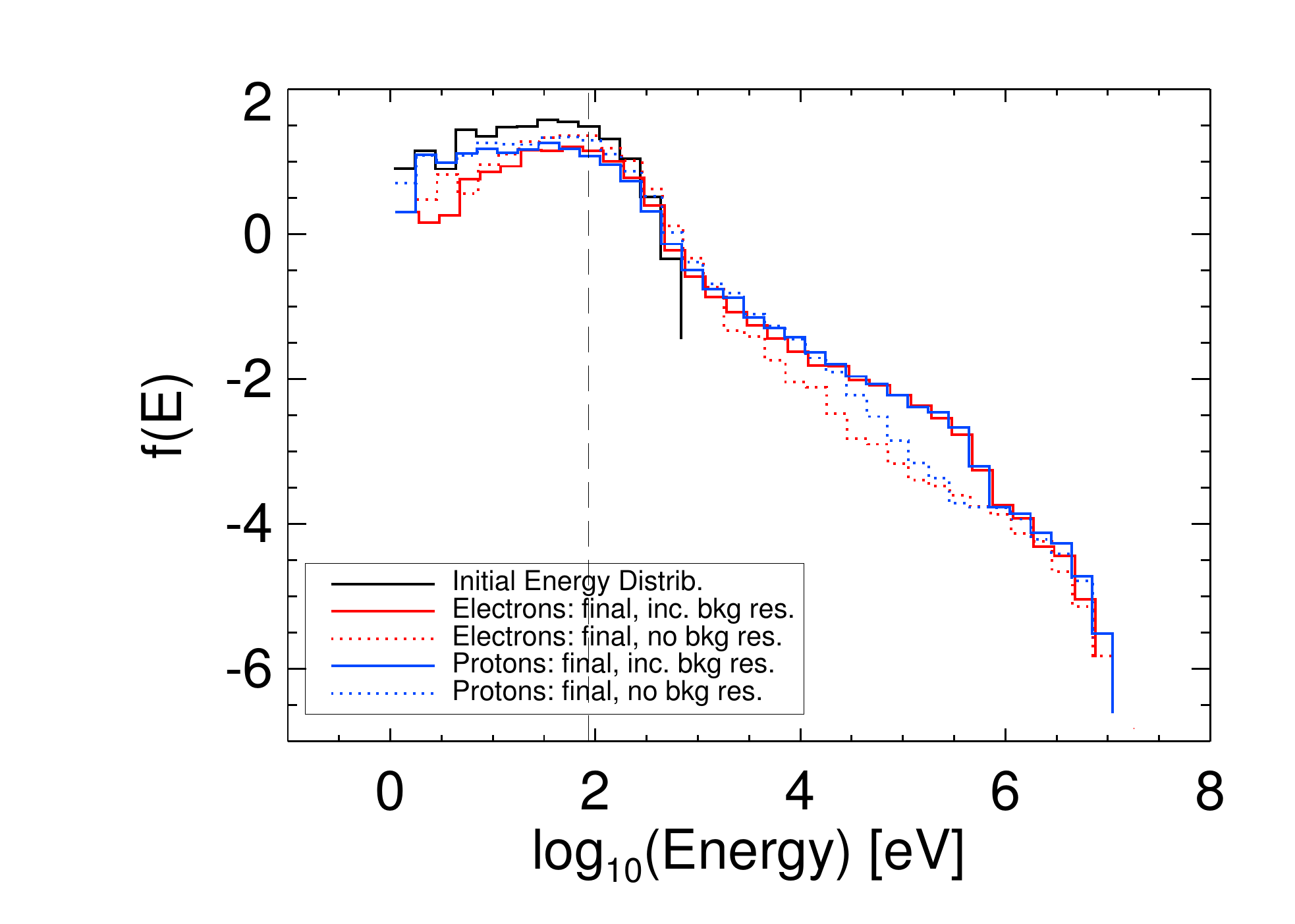}}} 
 \end{minipage}
  \caption{Single thread destabilises, Phase~2. \protect\subref{subfig:1Lt1pos} shows final positions and energies of proton orbits (for the case where orbit calculations omit background resistive effects) together with purple isosurface(s) of critical current and interpolated magnetic field lines (grey). Each final position is indicated by an orb, and coloured according to energy (for key, see colour bar). \protect\subref{subfig:1Lt1sp} \new{shows the initial and final energy distributions of both electron (red) and proton (blue) orbits where background resistivity is omitted (dashed histograms) or included (solid histograms). An example of the initial (Maxwellian) energy distribution is also included (black solid histogram)}.}
 \label{fig:oneloopt1}
\end{figure*}

\new{Figure~\ref{fig:oneloopt1} illustrates the final positions of proton orbits in three dimensions, colour-coded according to energy (Fig.~\ref{subfig:1Lt1pos}) together with the energy spectra of electron and proton orbits which include or omit background resistive effects (Fig.~\ref{subfig:1Lt1sp}). Orbit calculations in Phase~2 are based upon five MHD snapshots beginning at $t=60\tau_A$.

The kink instability onset occurs in the left-hand flux tube during Phase~2, resulting in the helical current sheet in Fig.~\ref{subfig:1Lt1pos}. The electric field generated by the resulting reconnection in this phase alters many orbit characteristics (previously dominated by the uniform resistive effects). When $\eta_{\rm{bkg}}$ is omitted, orbits in Fig.~\ref{subfig:1Lt1pos} form beams of energised protons accelerated towards both top and bottom boundaries in the left-hand flux tube alone. The peak electron energy reaches $22\unit{MeV}$, while the equivalent proton energy reaches $10.5\unit{MeV}$, caused solely by anomalous resistivity acting upon current $>j_{\rm{crit}}$.

The energy distributions in Fig.~\ref{subfig:1Lt1sp} reveal that orbit energisation is much more similar in cases where background resistivity is included (solid histograms) and omitted (dashed histogram), both for electrons and protons, compared to Phase~1. One notable difference between the two cases is that background resistivity increases the number of orbits with energies between $1\rm{keV}-1\rm{MeV}$ compared to anomalous resistivity alone.}

\subsection{Phase 3: excess current dissipated}\label{subsec:p3}
\begin{figure*}[t]
   \begin{minipage}[b]{0.48\textwidth}
    \subfloat[Proton final positions, energies \& field structure ($\eta_{\rm bkg}=0$)]{\label{subfig:1Lt2pos}
    \resizebox{\textwidth}{!}{\includegraphics{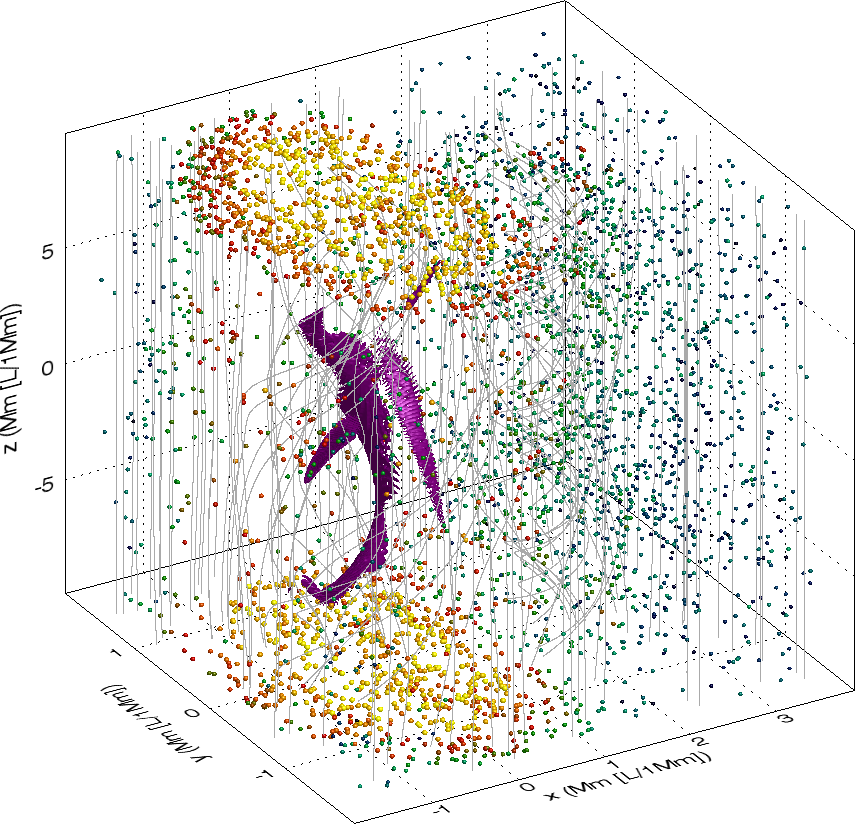}}}
   \end{minipage}
  \begin{minipage}[b]{0.45\textwidth}
  \resizebox{\textwidth}{!}{\includegraphics{AA31915f5x.png}}\\
  \subfloat[Energy distributions]{\label{subfig:1Lt2sp}\resizebox{\textwidth}{!}{\includegraphics[clip=true, trim=55 15 30 20]{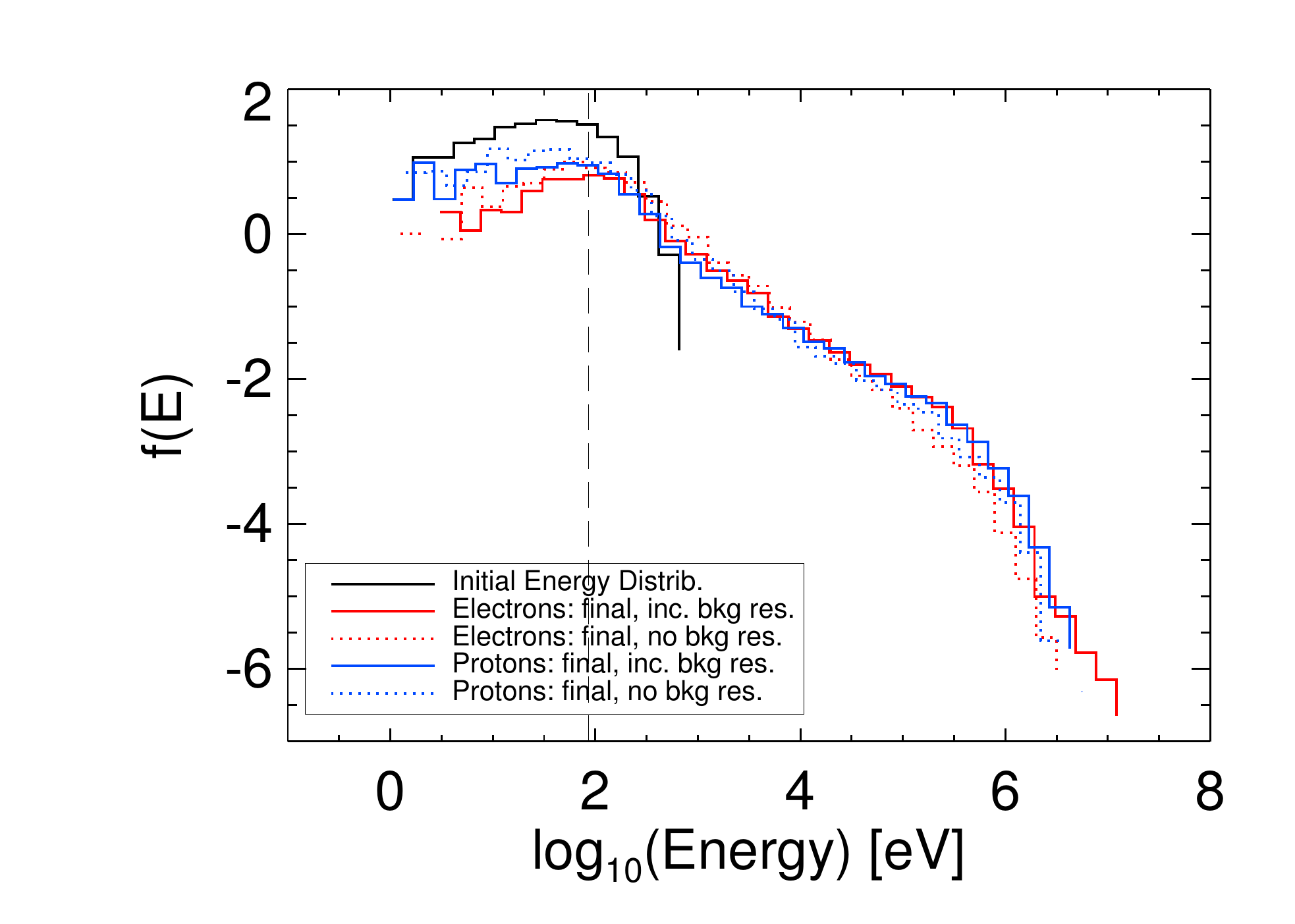}}} 
 \end{minipage}
  \caption{Single thread destabilises, Phase~3. \protect\subref{subfig:1Lt2pos} shows final positions and energies of protons (for the case where background resistivity is omitted from every orbit calculation) together with purple isosurface(s) of critical current and interpolated magnetic field lines (grey). The key to orbit energies is given in the colour bar displayed above \protect\subref{subfig:1Lt2sp}, \new{where the initial and final energy distributions of both electrons (red) and protons (blue) are displayed, for cases where background resistivity is omitted (dashed) or included (solid lines), while the black solid histogram shows an example of the initial energy distribution}.}
 \label{fig:oneloopt2}
\end{figure*}

The final phase occurs as the experiment nears its end. Orbit calculations which sample this phase use five MHD snapshots beginning at $t=110\tau_A$, with results presented in Fig.~\ref{fig:oneloopt2}. Once again, we present only the final positions of proton orbits which omit $\eta_{\rm{bkg}}$ in Fig.~\ref{subfig:1Lt2pos}, but compare the final energy spectra of electron and proton orbit calculations (as red and blue histograms) which include (solid lines) and omit background resistive effects (dashed) together in Fig.~\ref{subfig:1Lt2sp}.

Only a thin fragment of current remains above the critical value in Fig.~\ref{subfig:1Lt2pos}, near the centre of the left-hand flux tube remnants. While the current sheet itself is much smaller and thinner than in previous snapshots (suggesting that the impact sites may narrow), the remains of the tangled (reconnected) magnetic field allows orbits along many different field lines to access this acceleration region. Thus the final positions at the top and bottom boundaries widen (in Fig.~\ref{subfig:1Lt2pos}) compared to the final positions in previous phases. This expansion of the region occupied by energetic particles \citep[as noted by][]{paper:GordovskyyBrowning2011} is associated with reconnection of the twisted loop field lines with the ambient untwisted field. The peak electron and proton orbit energies drop to $4.2\unit{MeV}$ and $7.4\unit{MeV}$ respectively during this phase.

The energy spectra presented in Fig.~\ref{subfig:1Lt2sp} reveal a much closer match between resistivity profiles for both electron and proton orbits during Phase~3. Anomalous resistive effects dominate (as in Phase 2). \new{In all three Phases, a cross-over occurs close to $1\unit{keV}$; above this value, the spectra of orbits which incorporate $\eta_{\rm{bkg}}$ typically contain more orbits than calculations where it is omitted (and hence solid histograms lie above corresponding dashed cases above $1\unit{keV}$). With a limited total number of orbits, fewer orbits are therefore found at lower energies, and hence fewer orbits yield energies below $1\unit{keV}$ when including background resistive effects (causing the dashed histograms to typically lie above the solid histograms below this value). However, the differences between the cases are minor, compared to previous Phases.}

Since the background resistivity used in our simulations is greater than the coronal value, including this in the electric field when calculating particle orbit trajectories arguably significantly over-estimates particle acceleration efficiency and indeed (as in Phase 1) unrealistically predicts significant particle energisation in the absence of any reconnection. On the other hand, the anomalous resistivity within the current sheets represents the resistivity which is likely to be driven by plasma micro-instabilities, and gives a clear link between particle acceleration and electric fields associated with reconnection. Thus, in the following \citep[and similar to][]{paper:Gordovskyyetal2013,paper:Gordovskyyetal2014,paper:Gordovskyyetal2016,paper:Pintoetal2016}, we use only anomalous resistivity in calculating particle trajectories.

\section{Particle trajectories and energisation for interacting loops}\label{sec:twounstable}
\begin{figure*}[t]
 \centering
 \subfloat[$t=0\tau_{A}$]{\label{subfig:2Lt0B}\resizebox{0.32\textwidth}{!}{\includegraphics[clip=true, trim=110 100 110 190]{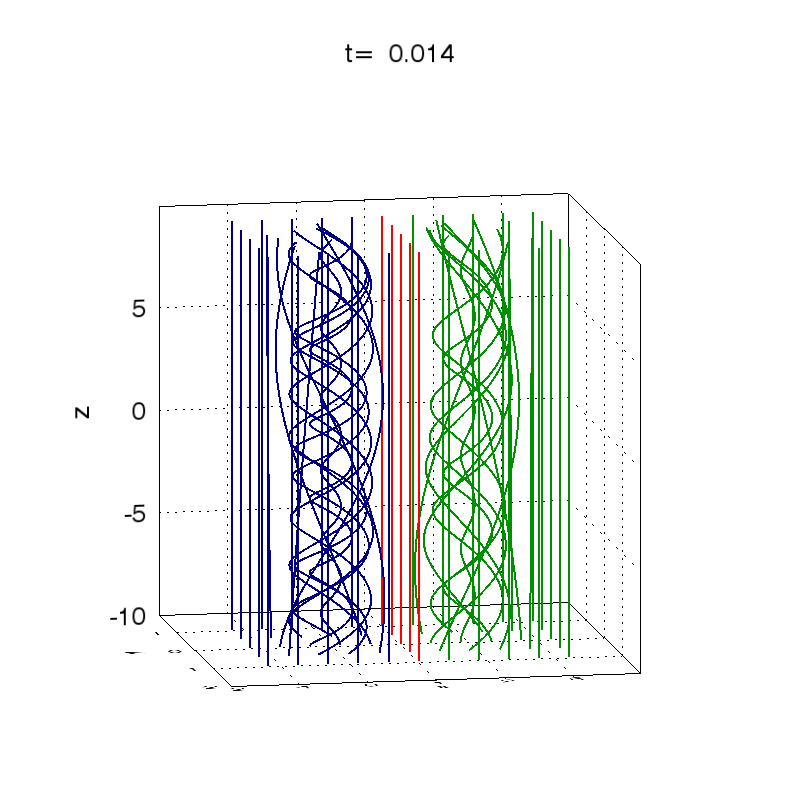}}}
 \subfloat[$t=80\tau_{A}$]{\label{subfig:2Lt1B}\resizebox{0.32\textwidth}{!}{\includegraphics[clip=true, trim=110 100 110 190]{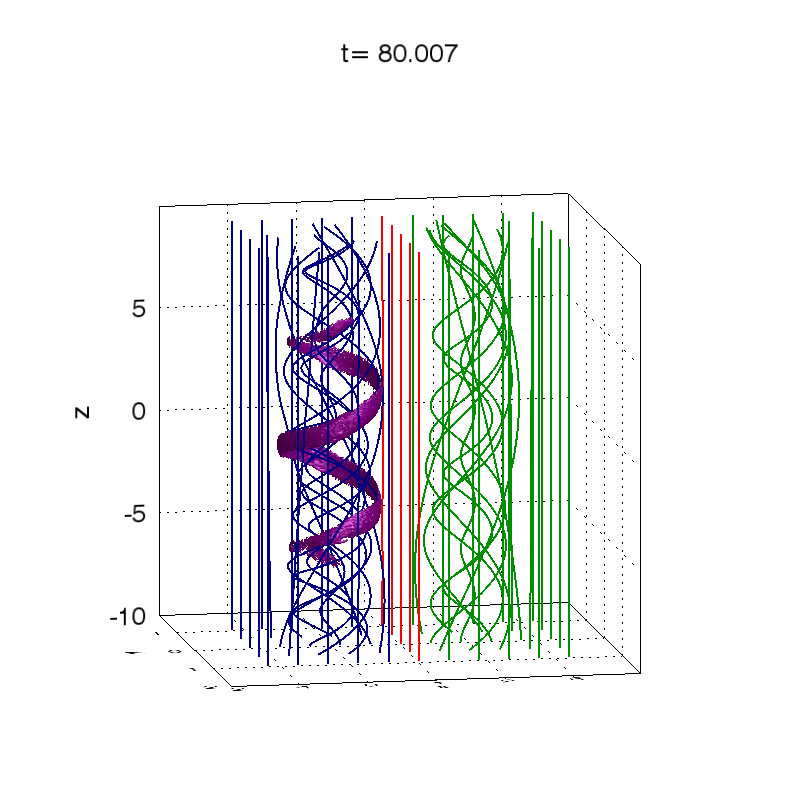}}} 
 \subfloat[$t=180\tau_{A}$]{\label{subfig:2Lt2B}\resizebox{0.32\textwidth}{!}{\includegraphics[clip=true, trim=110 100 110 190]{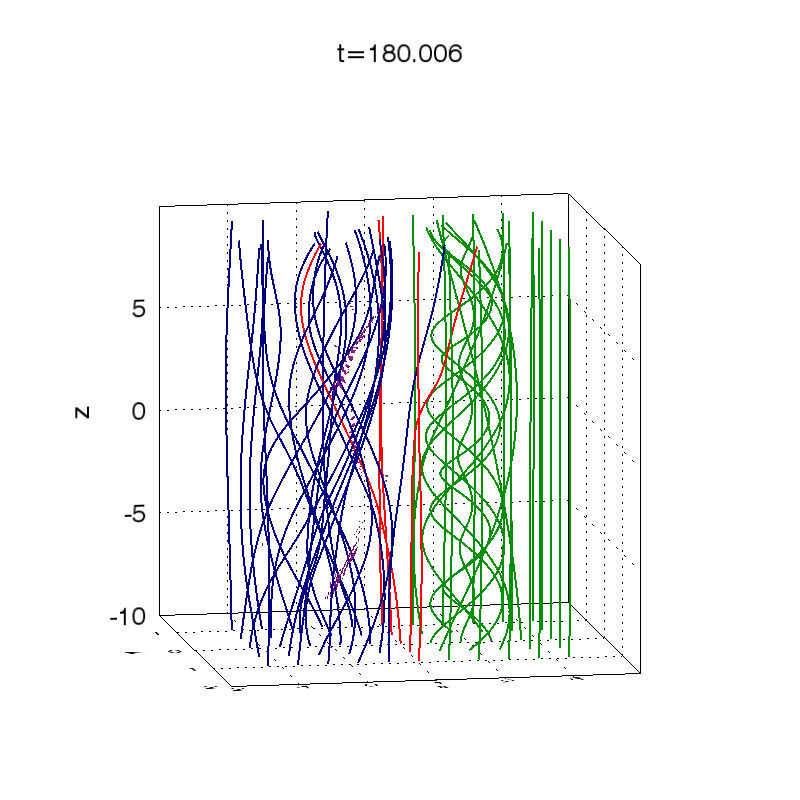}}}\\
  \subfloat[$t=220\tau_{A}$]{\label{subfig:2Lt3B}\resizebox{0.32\textwidth}{!}{\includegraphics[clip=true, trim=110 100 110 190]{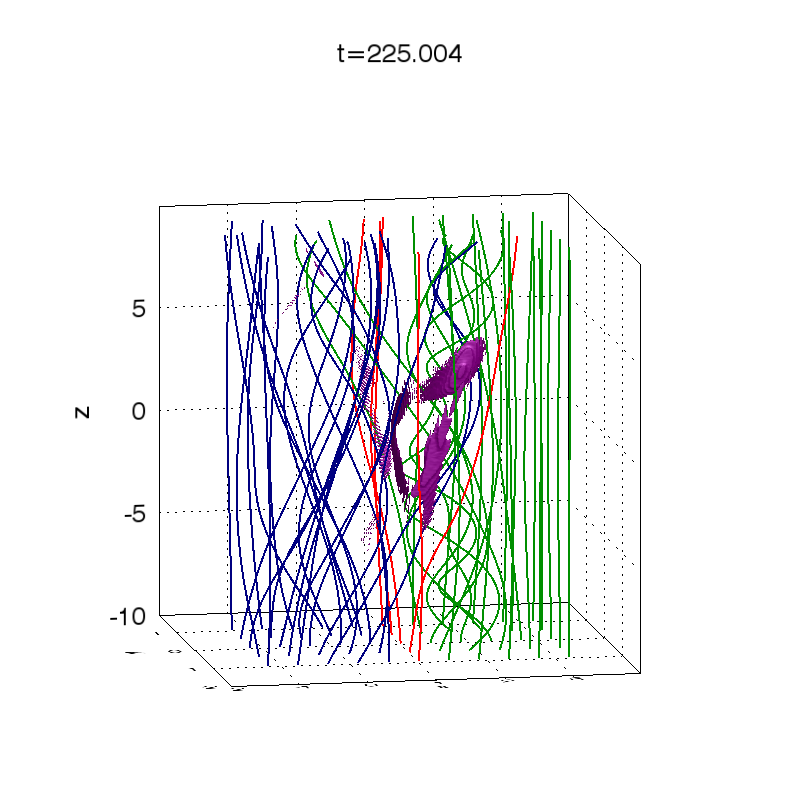}}}
 \subfloat[$t=250\tau_{A}$]{\label{subfig:2Lt4B}\resizebox{0.32\textwidth}{!}{\includegraphics[clip=true, trim=110 100 110 190]{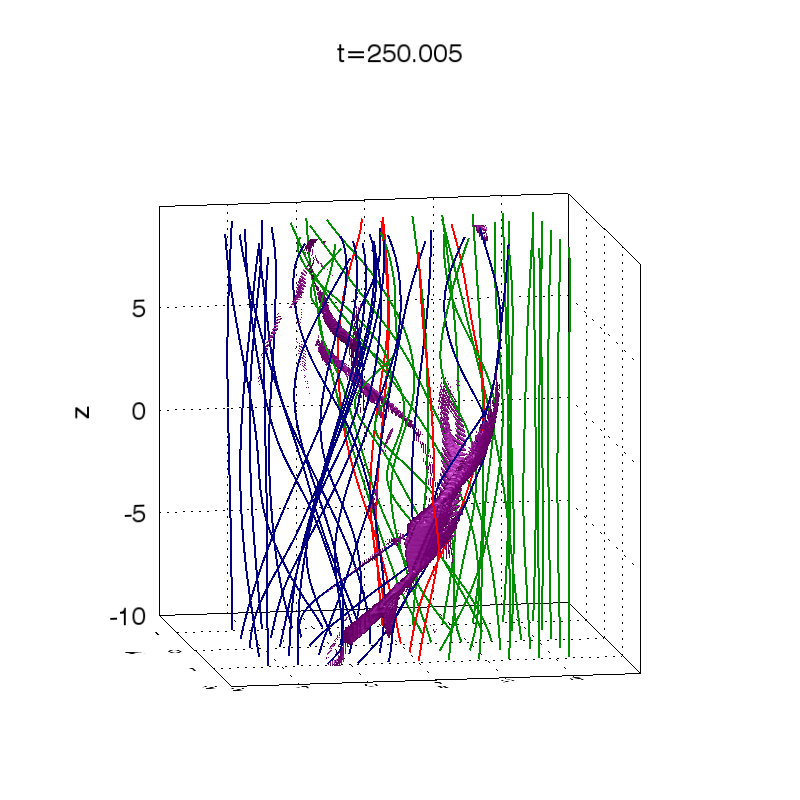}}}
 \subfloat[$t=300\tau_{A}$]{\label{subfig:2Lt5B}\resizebox{0.32\textwidth}{!}{\includegraphics[clip=true, trim=110 100 110 190]{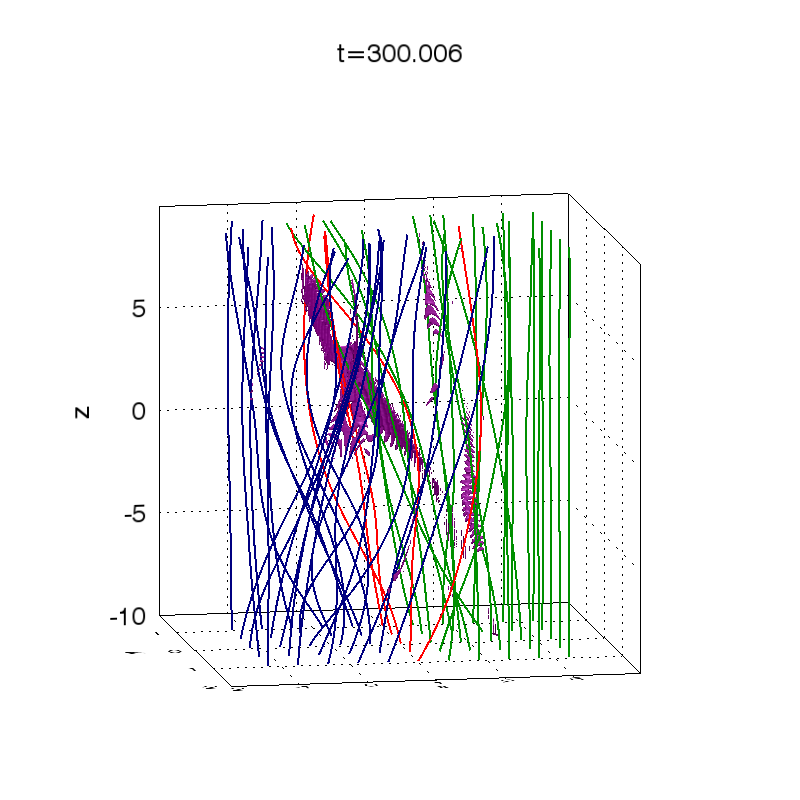}}}\\
  \caption{Interacting loops: Images show specific interpolated magnetic field lines (traced from the base of the simulation domain, at $\bar{z}=-10$ in non-dimensional units), colour-coded by location in $\bar{x}$, at different stages of the experiment. \new{Blue} field lines are traced from regions where $\bar{x}<-0.1$, \new{red} field lines are traced from $\bar{x}=0$ only, while \new{green} field lines are traced from $\bar{x}>0.1$. Thus, (prior to any reconnection) \new{blue} field lines are initially associated with the left-hand flux tube and \new{green} field lines with the right-hand tube, separated by \new{red} field lines. Purple isosurfaces (where present) indicate regions of current above the critical value.}
 \label{fig:twoloop_why}
\end{figure*}

We now consider the case where the left-hand flux tube destabilisation disrupts a second flux-tube, leading to a merger of the tubes. In doing so, we will use our earlier results as a guide, and omit uniform (background) resistive effects from our MHD and orbit calculations. 
\begin{figure}[t]
 \centering
  \resizebox{0.47\textwidth}{!}{\includegraphics[clip=true, trim=30 10 0 0]{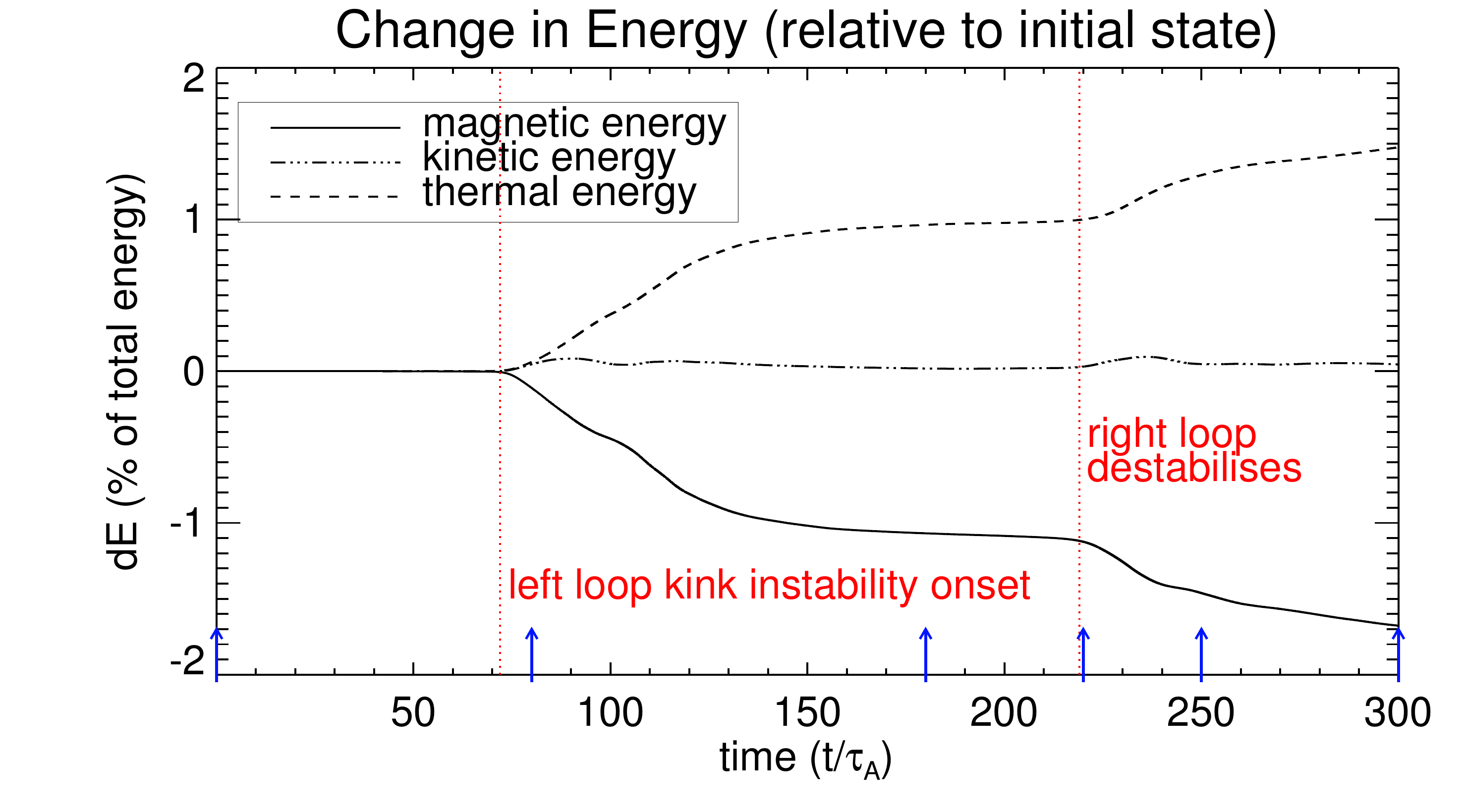}}
  \caption{Interacting loops: Energy partition as a function of time (in MHD simulation containing anomalous resistive effects only). \new{Red lines indicate approximate times when a loop destabilises, while blue arrows indicate times which are examined using particle orbit simulations, while} the key to individual energy components is given in the legend).}
 \label{fig:dEn2}
\end{figure}

As before, we begin with an outline of the MHD experiment. In Fig.~\ref{fig:twoloop_why}, we illustrate the evolution of magnetic field lines during the destabilisation of the first tube, and the disruption and merger with the second tube. The early evolution appears similar to that described in Section~\ref{sec:oneunstable}. The key difference can be seen in Fig.~\ref{subfig:2Lt1B}, with the appearance of the helical current sheet in the left hand tube. A comparison with Fig.~\ref{subfig:1Lt1B} \new{reveals that the centre of the helix of current at the midplane has moved from the side of the tube closest to its neighbour, and now lies on the opposite side of the tube in Fig.~\ref{subfig:2Lt1B}. In effect, the helical current sheet has been rotated by a factor of $\pi$ around the $z$-axis from the previous case, causing reconnection to initialise at different locations throughout the first tube.} The resulting flows and field line motion destabilise the `sheath' field between the flux tubes (Fig.~\ref{subfig:2Lt2B}). Crucially, the sheath field remained largely unaffected in the earlier experiment (see e.g. Fig.~\ref{subfig:1Lt2B}). In this experiment, the reconnection/removal of this sheath field ultimately leads to the disruption of the right-hand flux tube (\new{which begins in} Fig.~\ref{subfig:2Lt3B}, with \new{green} field lines now entering the region previously containing the left-hand flux tube), before the remaining currents further tangle the remaining tube remnants throughout the domain (Figs.~\ref{subfig:2Lt4B}-\ref{subfig:2Lt5B}). 

The energy components of the simulation also reflect this evolution, as shown in Fig.~\ref{fig:dEn2}. Several similarities are apparent when comparing Fig.~\ref{fig:dEn2} with the energy partition of the single loop case (Fig.~\ref{fig:dEn1}). Following an initial phase where no energy conversion takes place, magnetic energy begins to be largely converted into local heating at a rate which fluctuates over time, before returning to a near constant value. However, in this experiment, a second sharp loss of magnetic energy occurs. 

The first major loss of magnetic energy (at $t\approx72\tau_A$) marks the destabilisation of the left-hand tube, while the second major loss (at $t\approx219\tau_A$) marks the right-hand tube disruption and the merging of the loops. Additional bumps are visible on top of this general trend, for example at $t\approx110\tau_A$. To study these additional stages and features, we once again insert particles in the manner and amount described in the first experiment. Each set of orbits are calculated using five snapshots from different times throughout the MHD simulation, \new{focussing on specific times of interest identified by the blue arrows in Fig.~\ref{fig:dEn2}}. 
\begin{figure*}[t]
 \centering
 \subfloat[$t=0\tau_A$]{\label{subfig:2Lt0im}\resizebox{0.29\textwidth}{!}{\includegraphics[clip=true, trim=25 10 25 10]{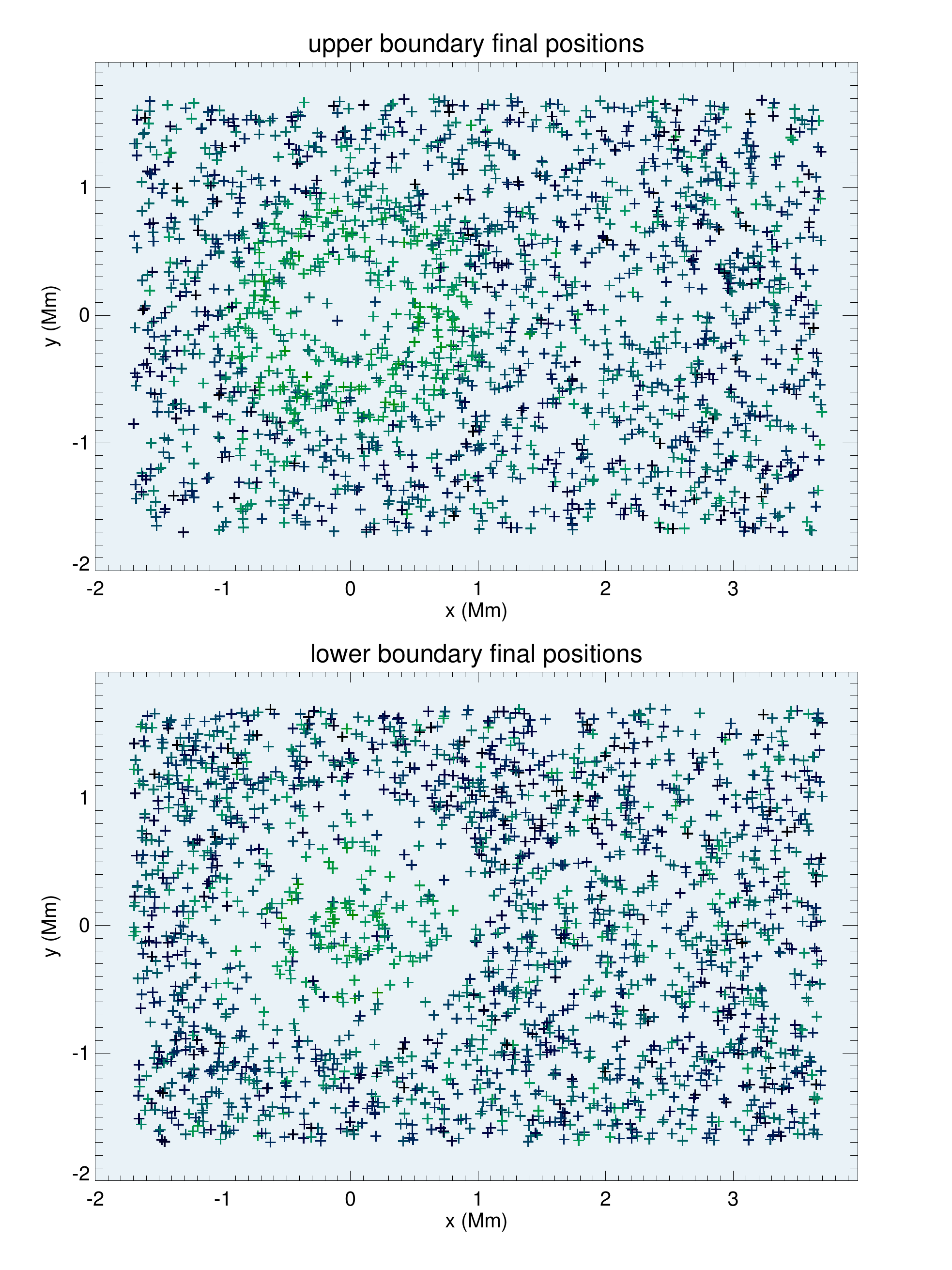}}}
 \subfloat[$t=90\tau_A$]{\label{subfig:2Lt1im}\resizebox{0.29\textwidth}{!}{\includegraphics[clip=true, trim=25 10 25 10]{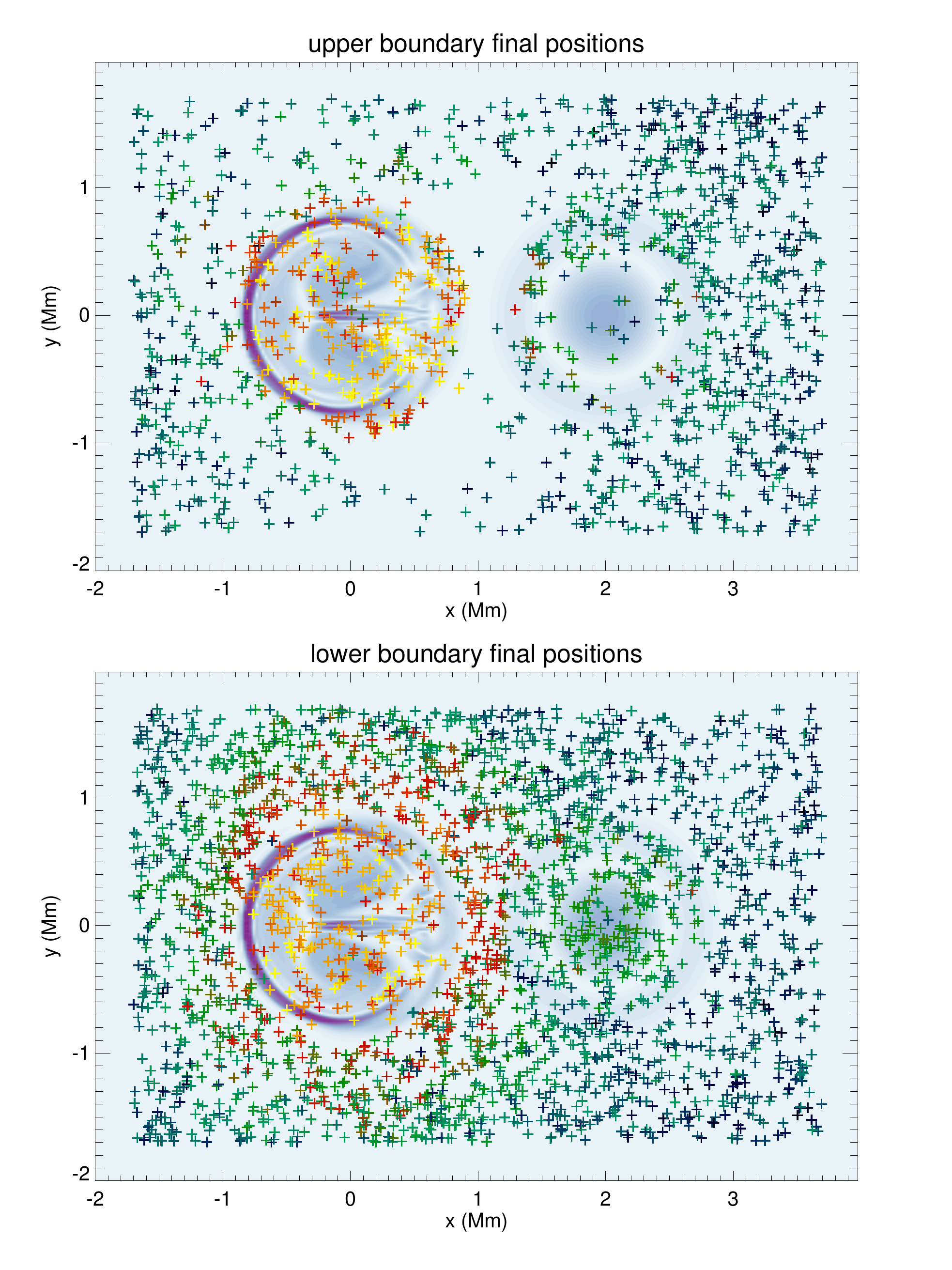}}}
 \subfloat[$t=115\tau_A$]{\label{subfig:2Lt2im}\resizebox{0.29\textwidth}{!}{\includegraphics[clip=true, trim=25 10 25 10]{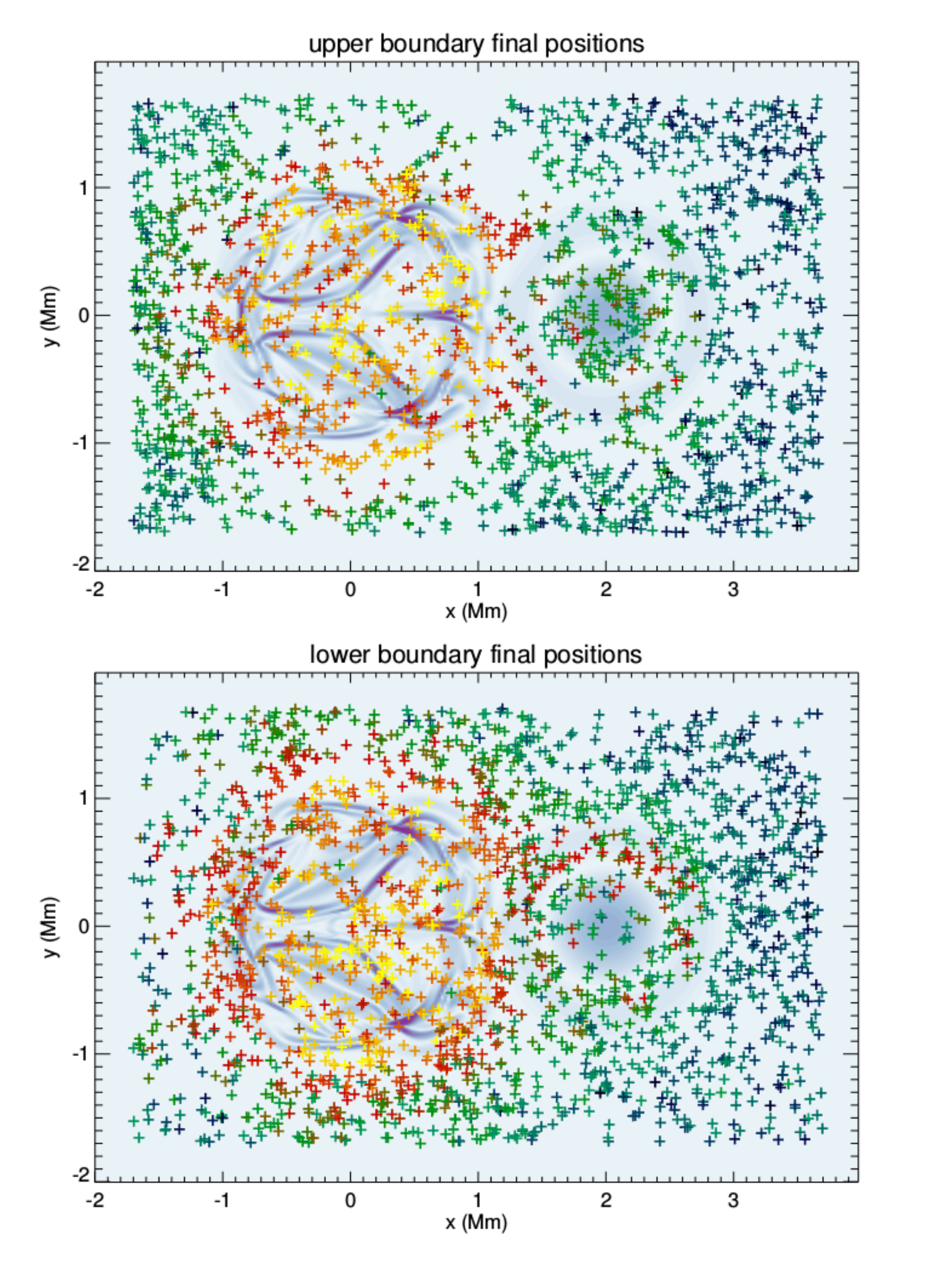}}}\hfill\resizebox{0.13\textwidth}{!}{\includegraphics{AA31915f9x1.png}}\\
 \subfloat[$t=190\tau_A$]{\label{subfig:2Lt3im}\resizebox{0.29\textwidth}{!}{\includegraphics[clip=true, trim=25 10 25 10]{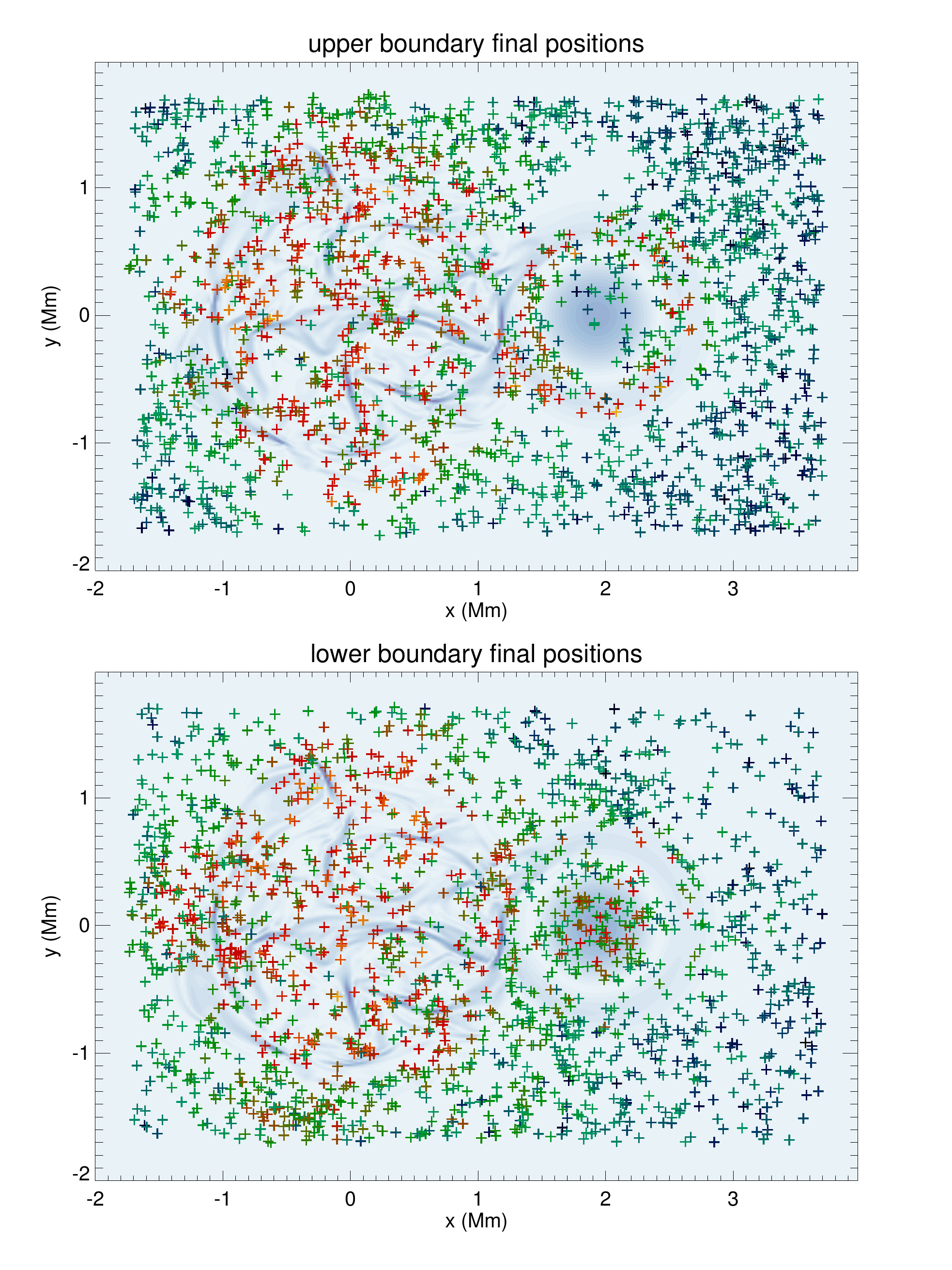}}}
 \subfloat[$t=235\tau_A$]{\label{subfig:2Lt4im}\resizebox{0.29\textwidth}{!}{\includegraphics[clip=true, trim=25 10 25 10]{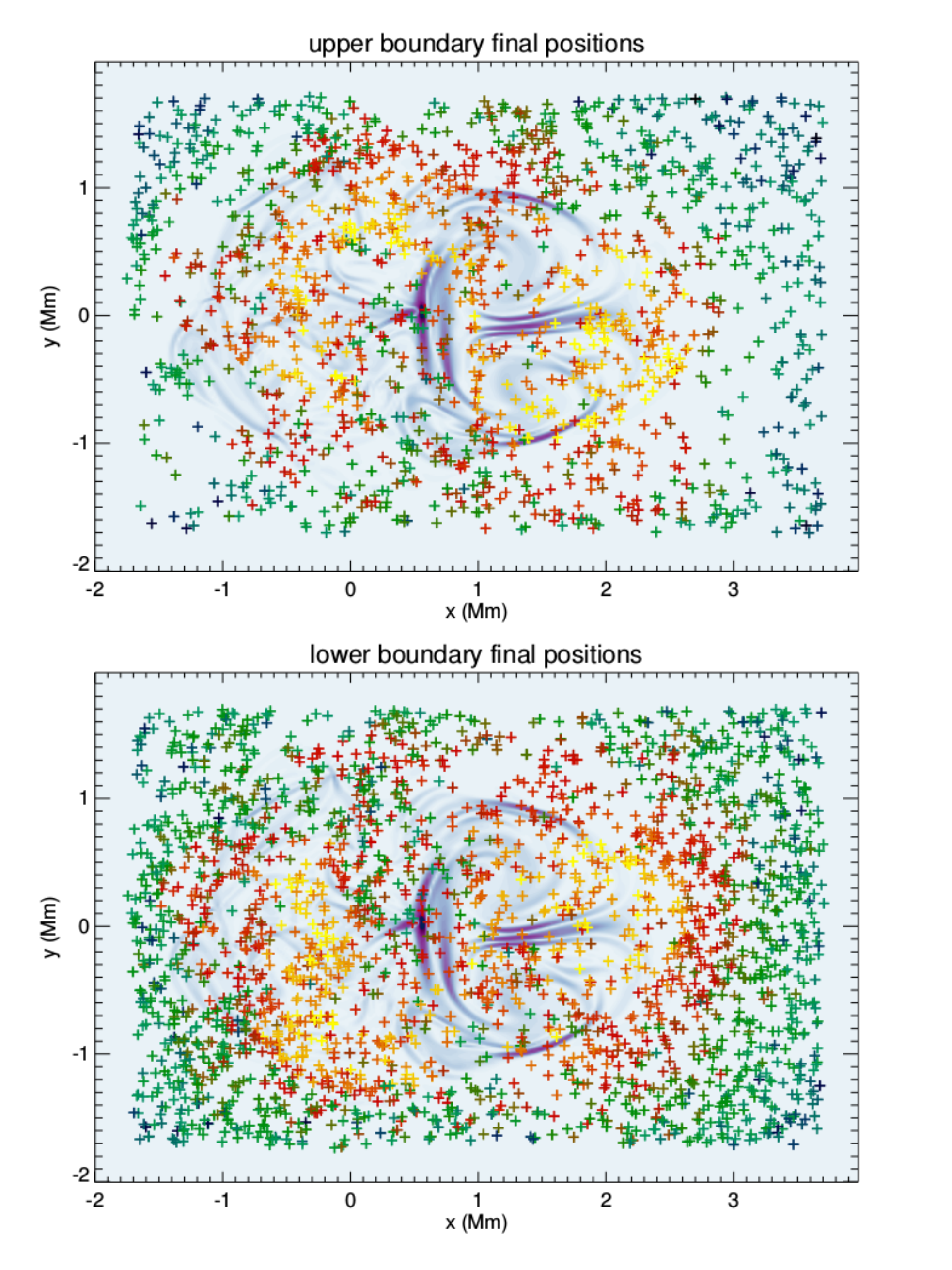}}}
 \subfloat[$t=270\tau_A$]{\label{subfig:2Lt5im}\resizebox{0.29\textwidth}{!}{\includegraphics[clip=true, trim=25 10 25 10]{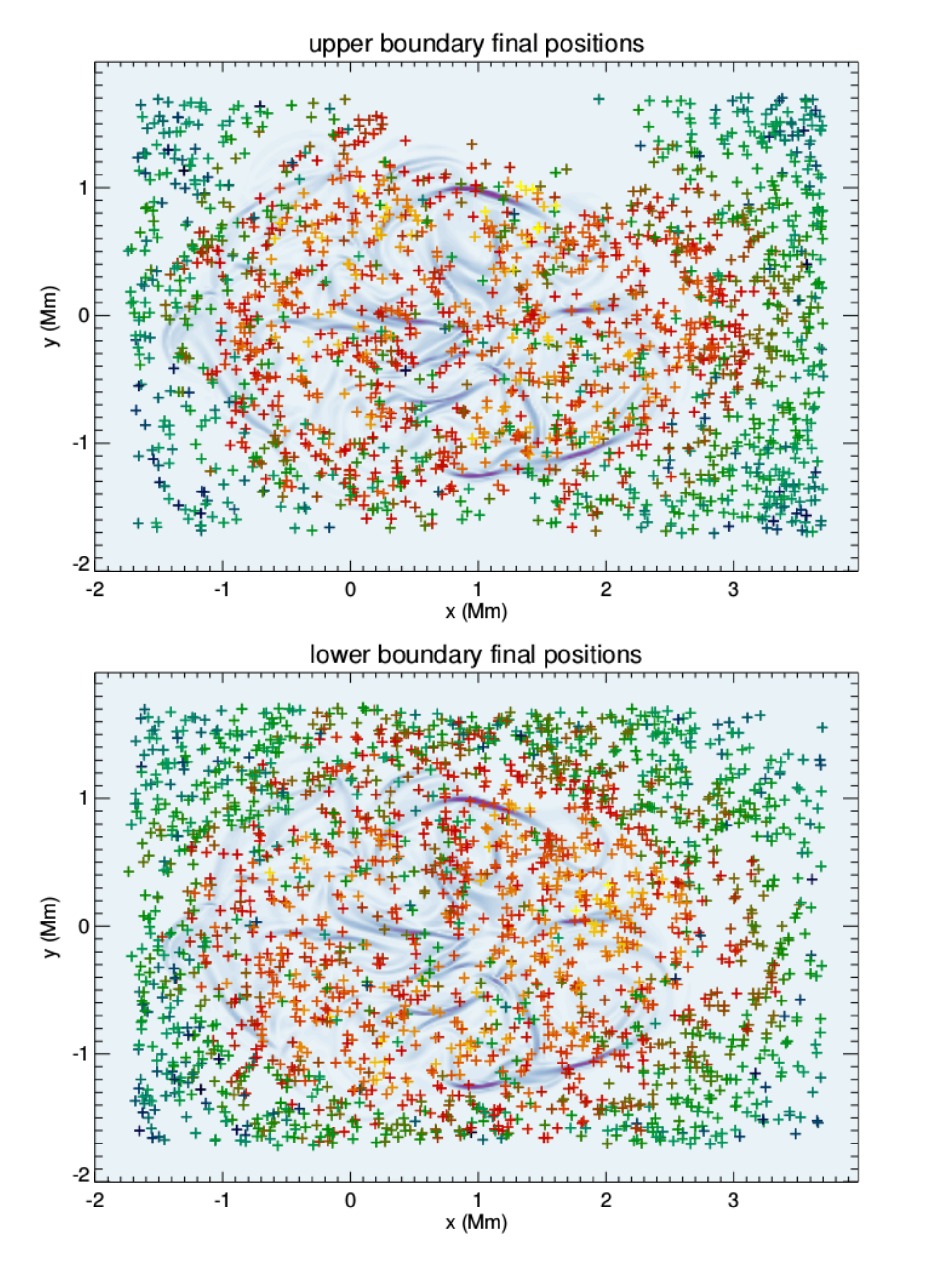}}}\hfill\resizebox{0.1\textwidth}{!}{\includegraphics{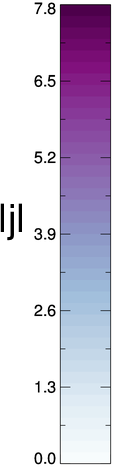}} 
  \caption{Interacting loops: Final electron orbit positions (colour-coded by orbit energy) passing through either the upper or lower boundary of the MHD simulations, using different snapshots of the MHD loop experiment in which two loops merge. The background colour indicates the current structure ($|j|$) in the midplane of the domain, at the time when the orbits are initialised (with the key to the background colour seen in the second colour bar, in non-dimensional units).}
 \label{fig:twoloops_impact}
\end{figure*}

The evolution of final electron orbit positions at each boundary over time can be seen in Fig.~\ref{fig:twoloops_impact}, with each location colour-coded according to the orbit energy upon arrival at the boundary. Figure~\ref{fig:twoloops_impact} reveals how the fragmentation and merger of the tubes affect particle acceleration at different stages of the experiment. Energised final positions prior to the onset of the kink instability can be seen in Fig.~\ref{subfig:2Lt0im}, and are well matched to those seen in Phase~1 of the single loop case (e.g. Fig.~\ref{subfig:1Lt0pos2}). 
The helical current sheet begins to form at approximately $t=70\tau_A$, and is well developed by $t=90\tau_A$. By this time, as shown in Fig.~\ref{subfig:2Lt1im}, the current sheet generates near-symmetrical beams of high energy electrons which reach both the upper and lower footpoints of the left-hand tube, surrounded by a halo-like structure of final positions at more moderate energies, particularly at the lower boundary. The energised locations widen by $t=115\tau_A$ (seen in Fig.~\ref{subfig:2Lt2im}), while the surrounding moderate energy gains are much more evenly spread between top and bottom boundaries.
The number of high energy orbits reaching either boundary decreases sharply by $t=190\tau_A$ (Fig.~\ref{subfig:2Lt3im}). This stage coincides with the rate of magnetic energy conversion to heat returning almost to zero (noting a lack of yellow symbols in Fig.~\ref{fig:dEn2}). Several \new{thermal energy} orbits reach the boundary during this stage, but sporadically and without association with any specific region/grouping. The onset of the disruption in the right-hand tube causes significant acceleration once again, as shown in Fig.~\ref{subfig:2Lt4im}, but crucially with signatures associated with the footpoints of both flux tubes. By the final stage of the experiment, in Fig.~\ref{subfig:2Lt5im}, high energy orbit locations are again \new{largely absent}, while moderate energy locations are spread widely over both boundaries.
\begin{figure*}[t]
 \centering
 \subfloat[$t=90{\tau_A}$]{\label{subfig:2Lt2pos}\resizebox{0.4\textwidth}{!}{\includegraphics{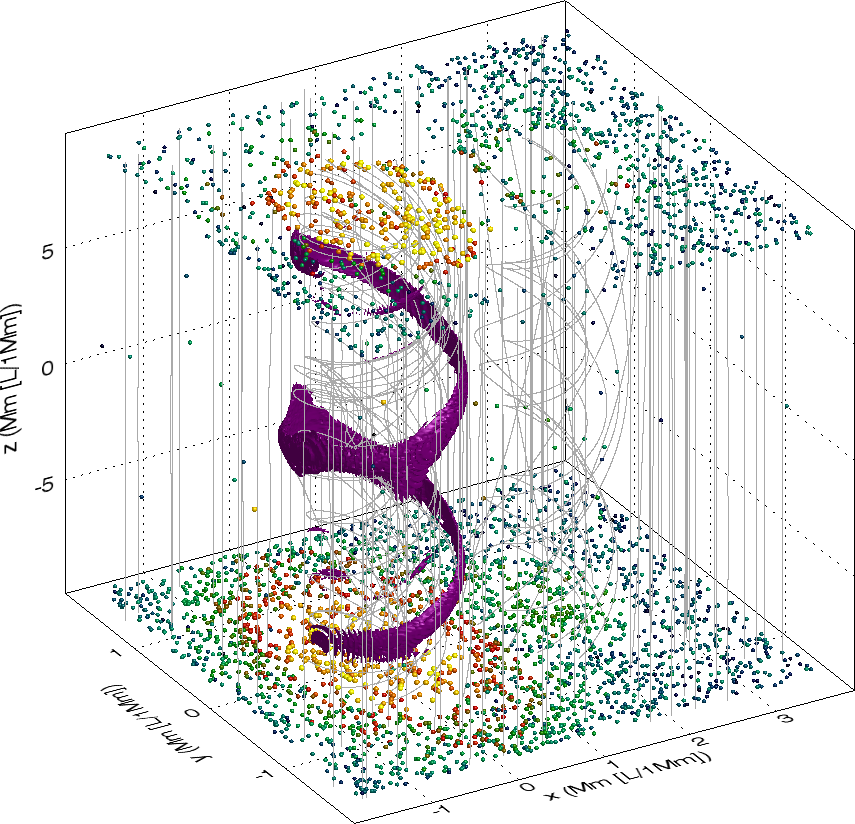}}}
 \subfloat[$t=115\tau_A$]{\label{subfig:2Lt3pos}\resizebox{0.4\textwidth}{!}{\includegraphics{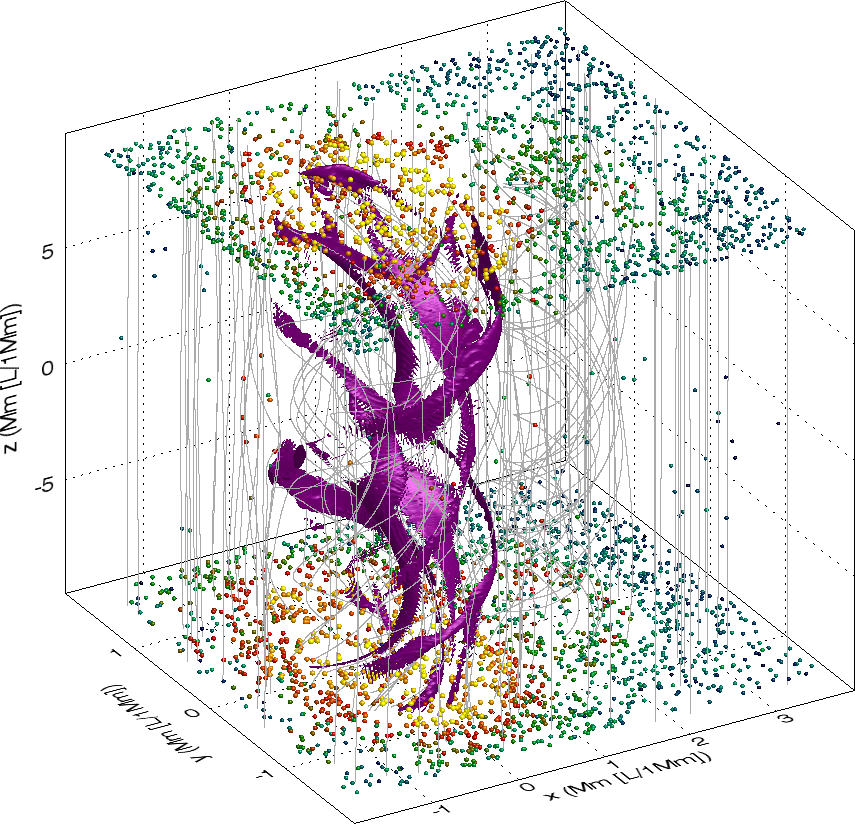}}}
 \hspace{0.01\textwidth}\resizebox{0.12\textwidth}{!}{\includegraphics{AA31915f9x1.png}}\\
 \subfloat[$t=190{\tau_A}$]{\label{subfig:2Lt4pos}\resizebox{0.4\textwidth}{!}{\includegraphics{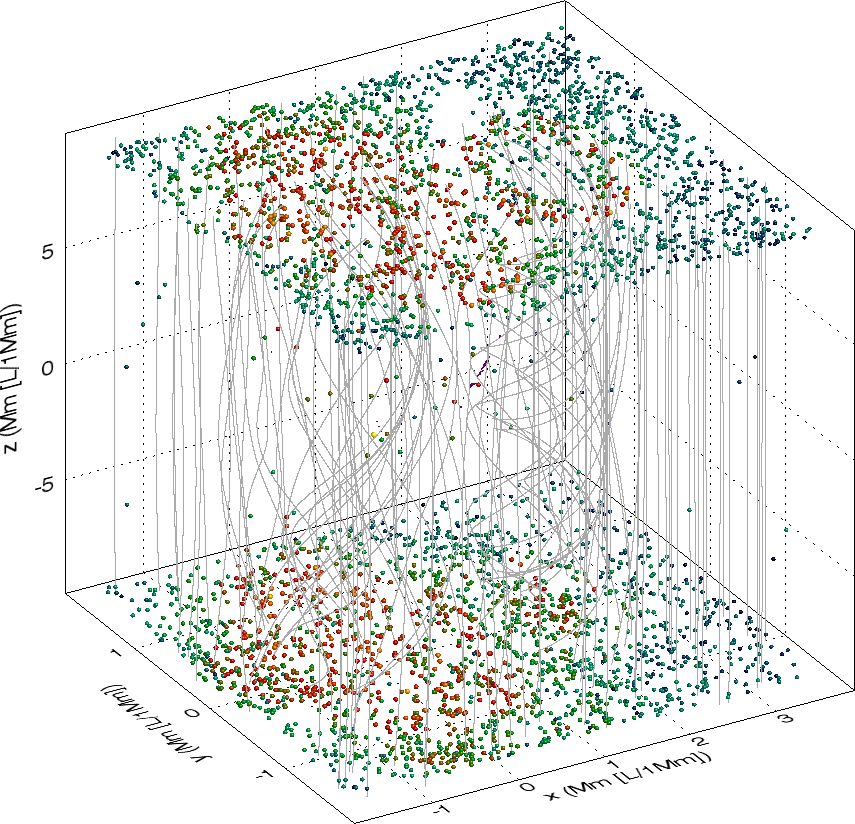}}}
 \subfloat[$t=235\tau_A$]{\label{subfig:2Lt5pos}\resizebox{0.4\textwidth}{!}{\includegraphics{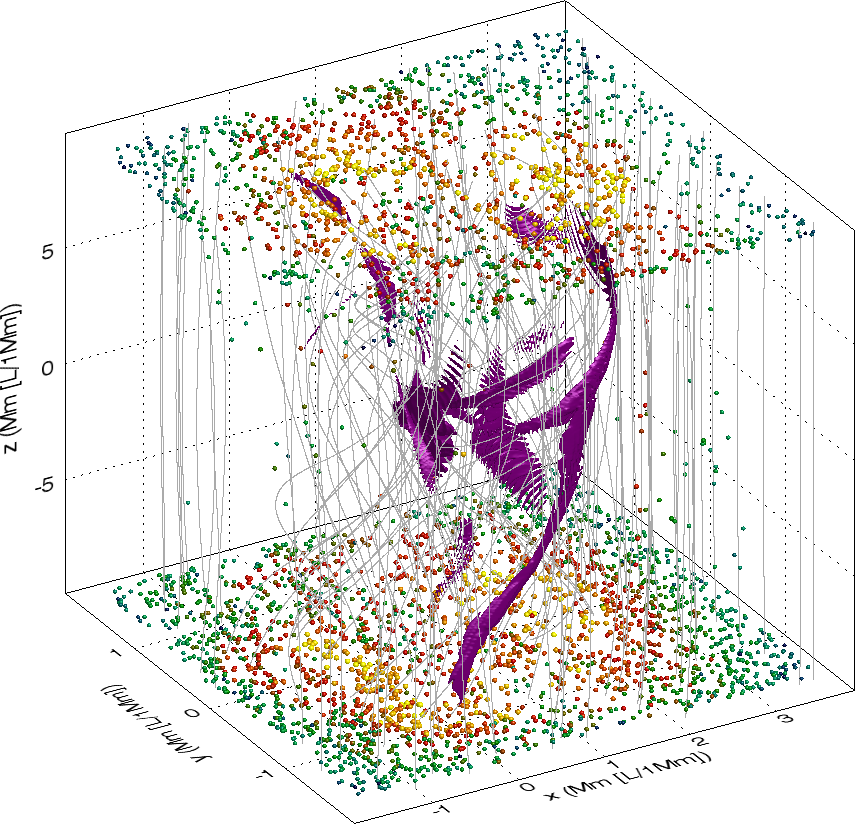}}}
 \hspace{0.01\textwidth}\resizebox{0.12\textwidth}{!}{\includegraphics{AA31915f9x1.png}}\\
   \caption{Interacting loops: 3D map of final positions (coloured orbs) and energies of electron orbits initialised at two different snapshots of the MHD loop experiment (before and during stages when the loops merge). Isosurfaces of current above the critical current are included (where present), while key to energy is given by the colour bar. }
 \label{fig:twoloops_pos}
\end{figure*}

The crucial difference between the merging loop case studied here and the single loop case studied in Sec.~\ref{sec:oneunstable} is the secondary disruption. To emphasise our primary findings, Fig.~\ref{fig:twoloops_pos} illustrates the difference in final positions and energies of electron orbits in three dimensions. These figures compare current isosurfaces and final positions associated with the left-hand loop destabilisation (Figs.~\ref{subfig:2Lt2pos} and~\ref{subfig:2Lt3pos}), and the phases prior to and after the right-hand loop destabilises (Figs.~\ref{subfig:2Lt4pos} and~\ref{subfig:2Lt5pos}). Few orbits are accelerated to any degree in between disruptions. Figure.~\ref{subfig:2Lt5pos} emphasises how the ribbons of current causes orbits with non-thermal energies to be associated with the former footpoints of both tubes. 
\begin{figure*}[t]
 \centering
 \subfloat[$t=0\tau_A$]{\label{subfig:2Lt0imP}\resizebox{0.29\textwidth}{!}{\includegraphics[clip=true, trim=25 10 25 10]{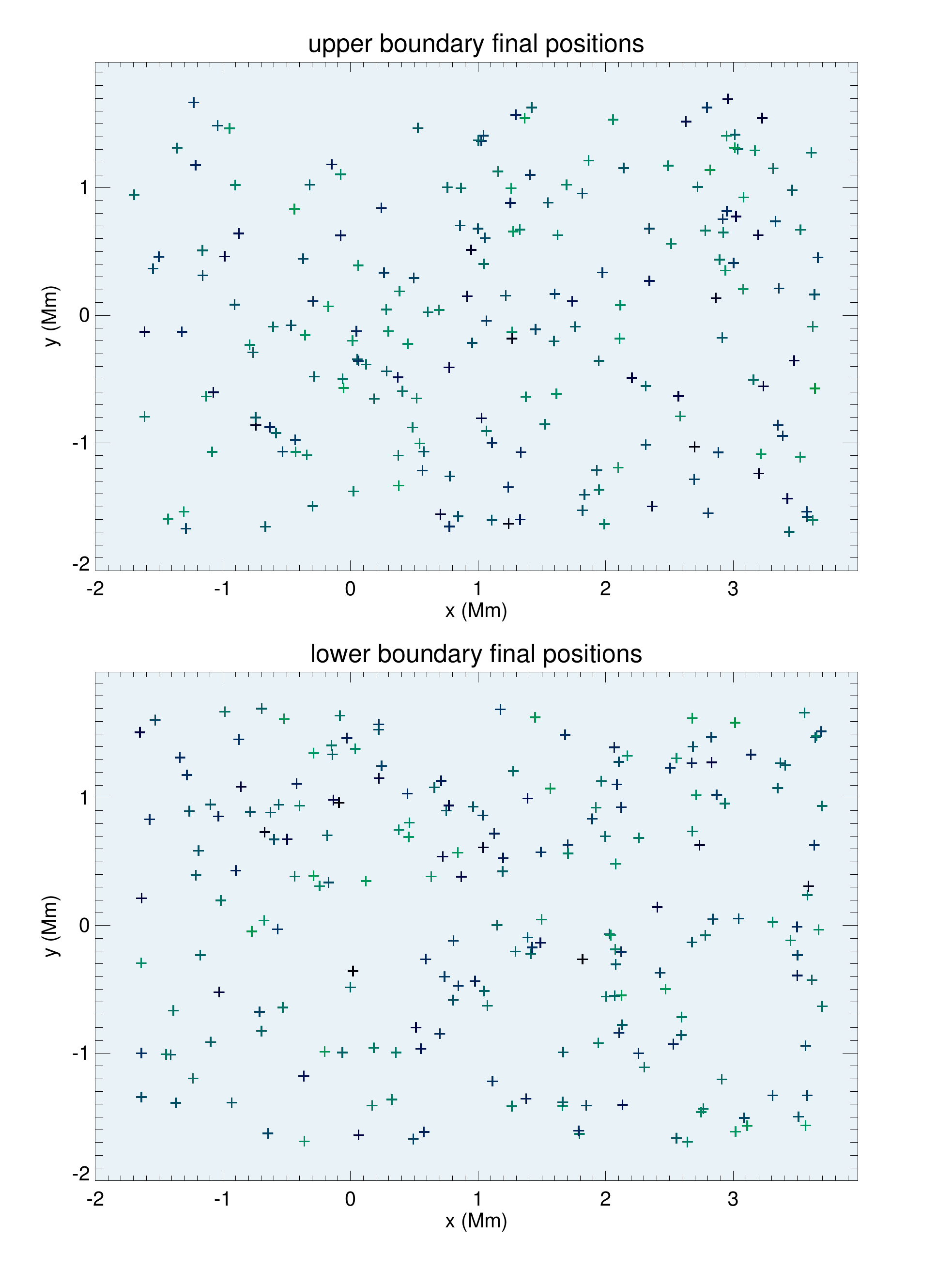}}}
 \subfloat[$t=90\tau_A$]{\label{subfig:2Lt1imP}\resizebox{0.29\textwidth}{!}{\includegraphics[clip=true, trim=25 10 25 10]{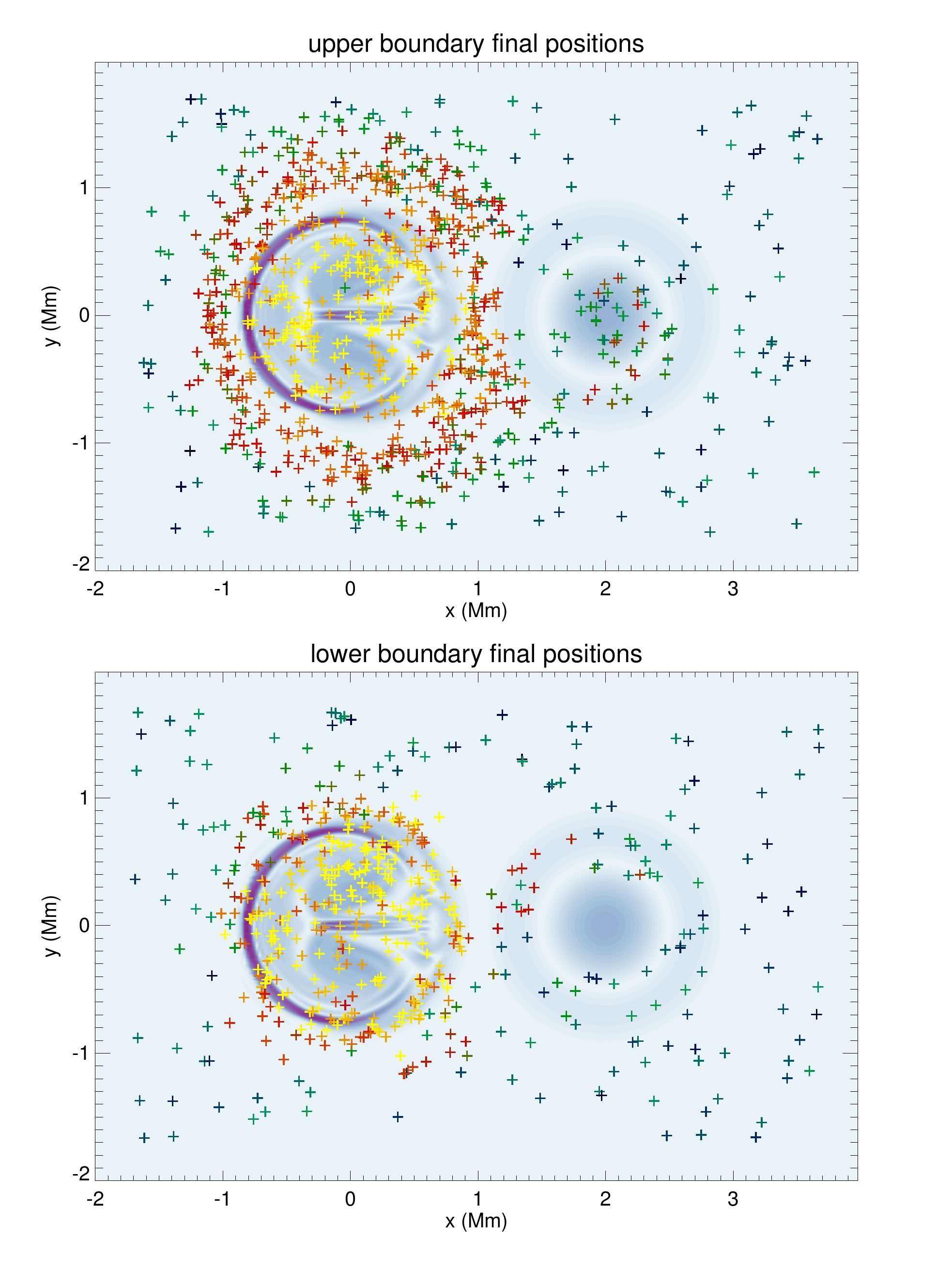}}}
 \subfloat[$t=115\tau_A$]{\label{subfig:2Lt2imP}\resizebox{0.29\textwidth}{!}{\includegraphics[clip=true, trim=25 10 25 10]{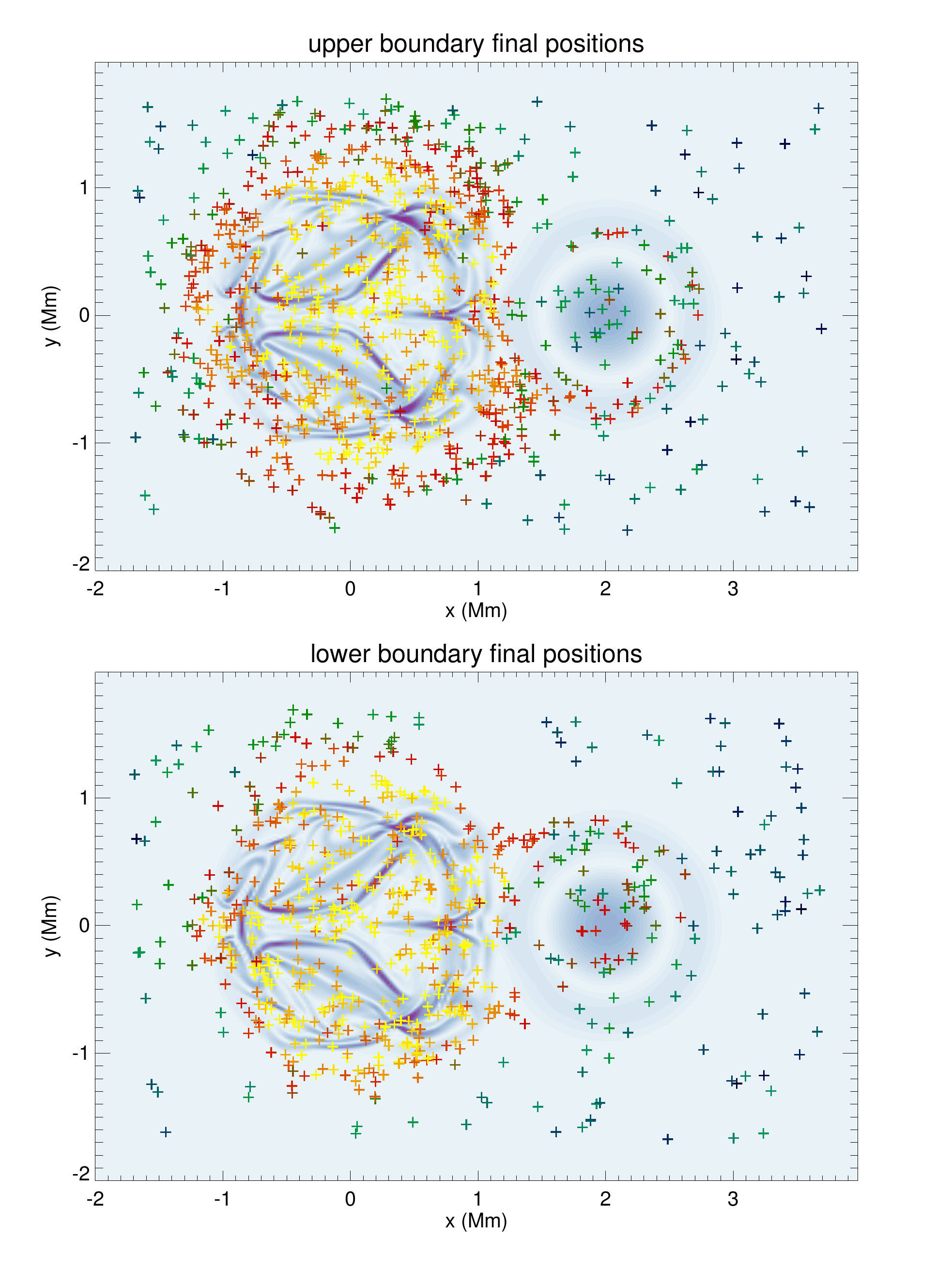}}}\hfill\resizebox{0.13\textwidth}{!}{\includegraphics{AA31915f9x1.png}}\\
 \subfloat[$t=190\tau_A$]{\label{subfig:2Lt3imP}\resizebox{0.29\textwidth}{!}{\includegraphics[clip=true, trim=25 10 25 10]{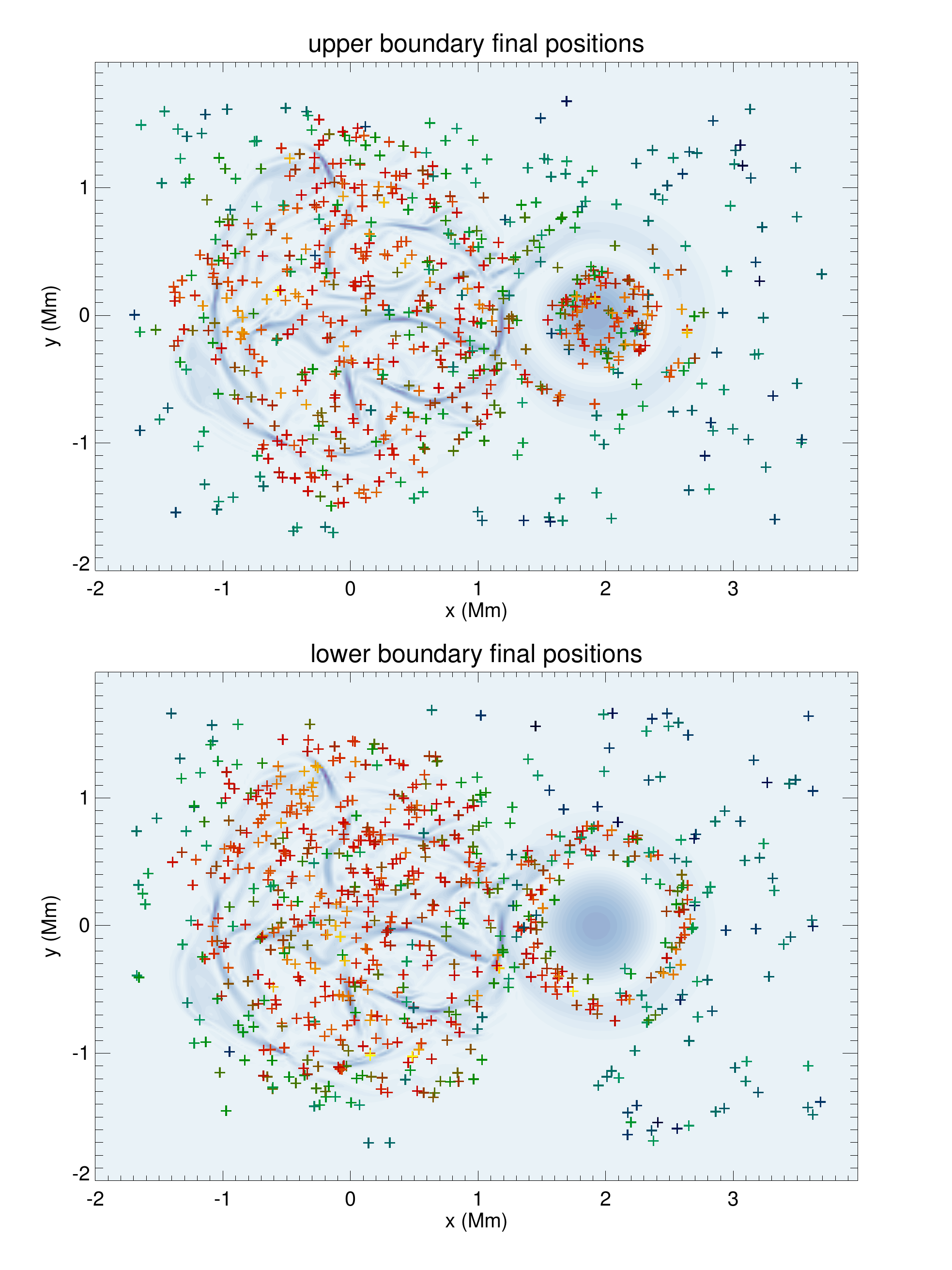}}}
 \subfloat[$t=235\tau_A$]{\label{subfig:2Lt4imP}\resizebox{0.29\textwidth}{!}{\includegraphics[clip=true, trim=25 10 25 10]{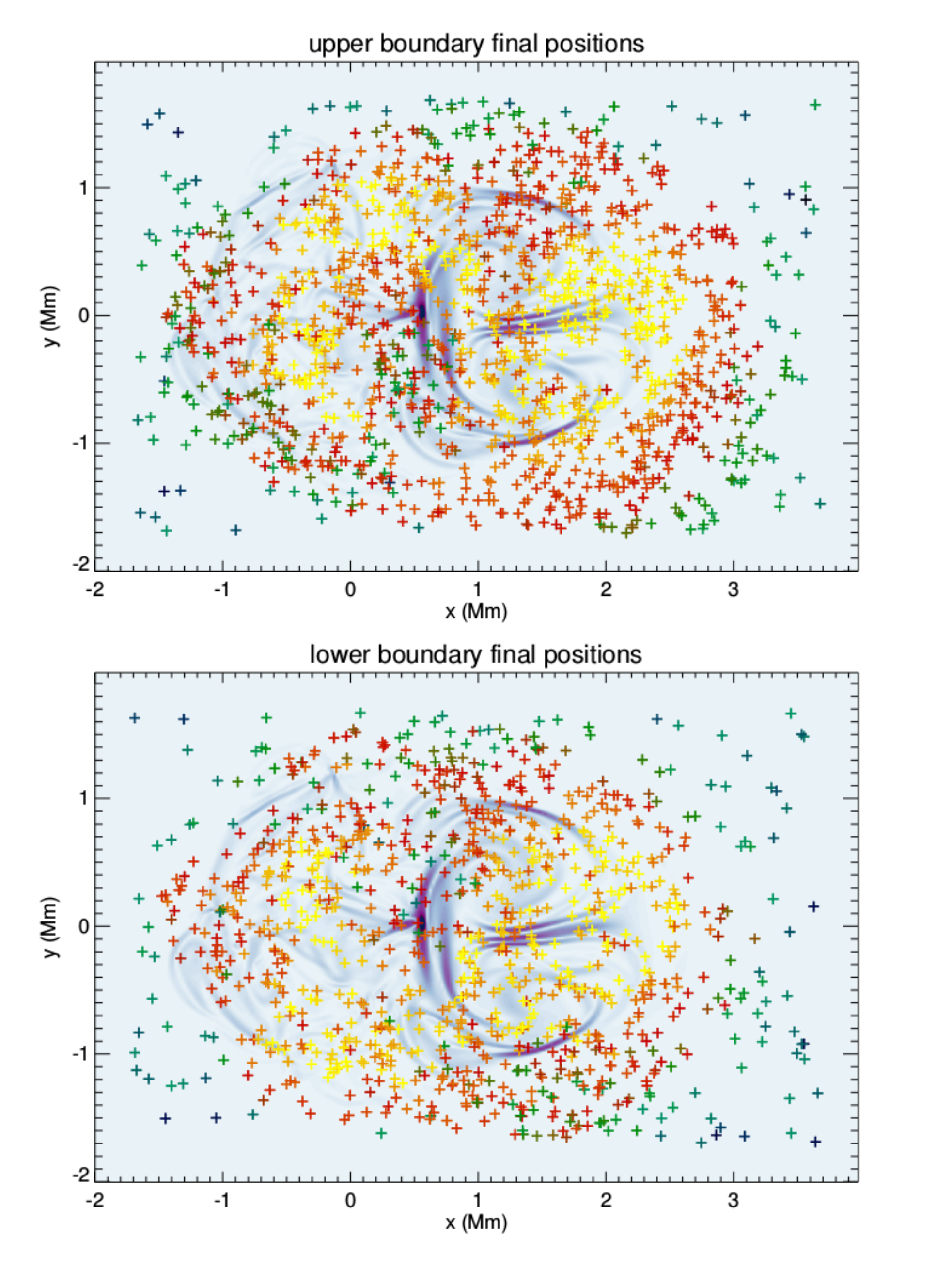}}}
 \subfloat[$t=270\tau_A$]{\label{subfig:2Lt5imP}\resizebox{0.29\textwidth}{!}{\includegraphics[clip=true, trim=25 10 25 10]{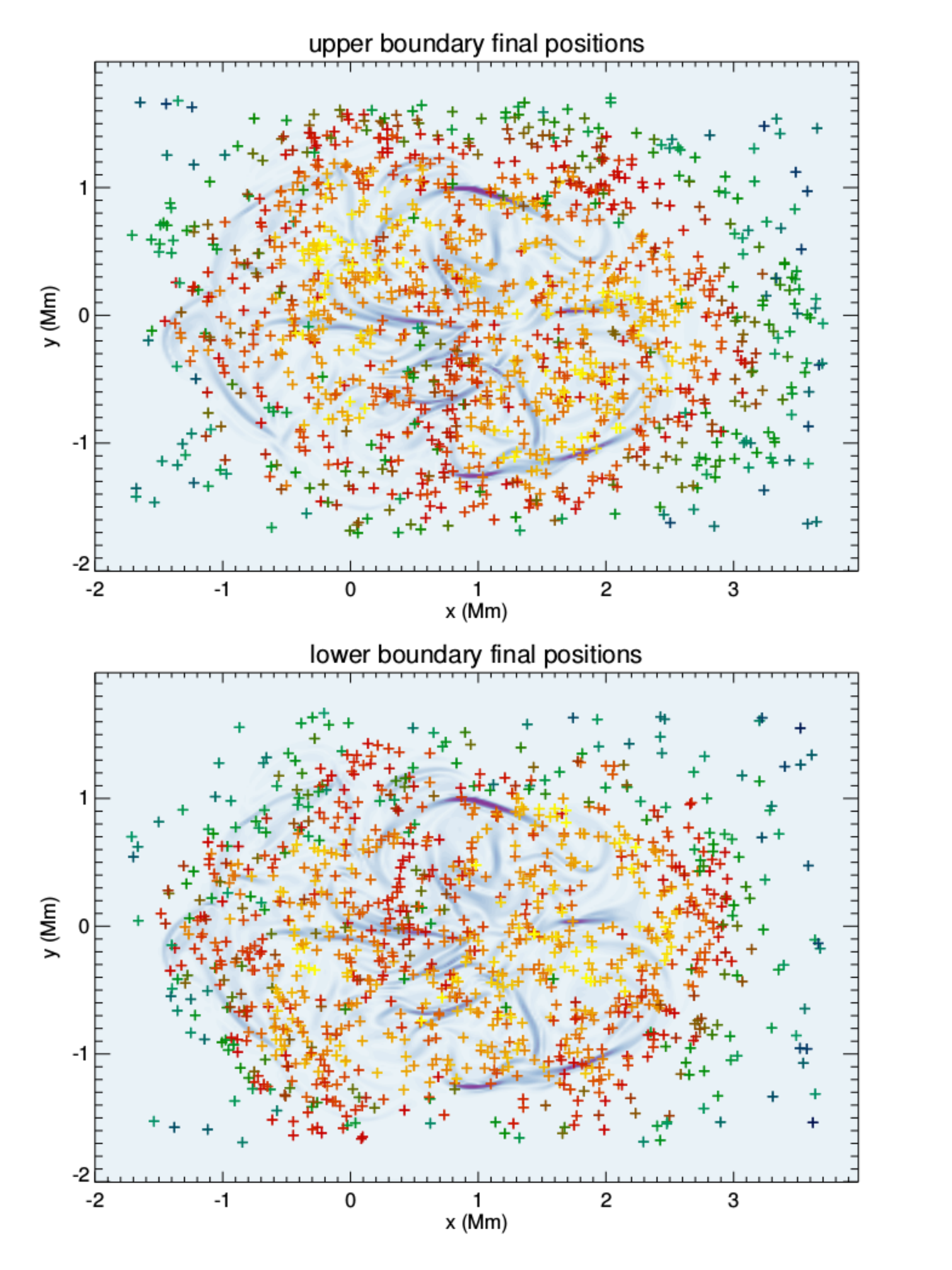}}}\hfill\resizebox{0.1\textwidth}{!}{\includegraphics{AA31915f9x2.png}} 
  \caption{Interacting loops: Final proton orbit positions (colour-coded by orbit energy) passing through either the upper or lower boundary of the MHD simulations, using different snapshots of the MHD loop experiment in which two loops merge. The background colour indicates the current structure ($|j|$) in the midplane of the domain, at the time when the orbits are initialised (with the key to the background colour seen in the second colour bar, in non-dimensional units).}
 \label{fig:twoloops_impactP}
\end{figure*}

For comparison, Fig.~\ref{fig:twoloops_impactP} illustrates the final proton orbit positions and energies, at identical stages to the electron orbits seen in Fig.~\ref{fig:twoloops_impact}. As with the bulk of the experiment, the proton impact sites largely bear out the findings of the electrons. Fewer protons reach the top and bottom boundaries in the same time compared to electrons, and hence Fig.~\ref{subfig:2Lt0imP} effectively shows random locations and energies without evidence of acceleration. The destabilisation of the left-hand tube causes widening beams of protons to reach both boundaries in Figs.~\ref{subfig:2Lt1imP} and~\ref{subfig:2Lt2imP}, before returning to a much quieter phase at in Fig.~\ref{subfig:2Lt3imP}. The destabilisation of the second loop sees the formation of near-symmetric beams of high energy orbits forming in Fig.~\ref{subfig:2Lt4imP} which stretch between the footpoint locations of both tubes. By the final stage, Fig.~\ref{subfig:2Lt5imP} shows that most energised final positions are clustered near the centre of the domain, without any visible distinction between flux tubes.
\begin{figure*}[t]
 \centering
  \subfloat[electron spectra]{\label{subfig:2Lspec}\resizebox{0.49\textwidth}{!}{\includegraphics[clip=true, trim=55 15 30 20]{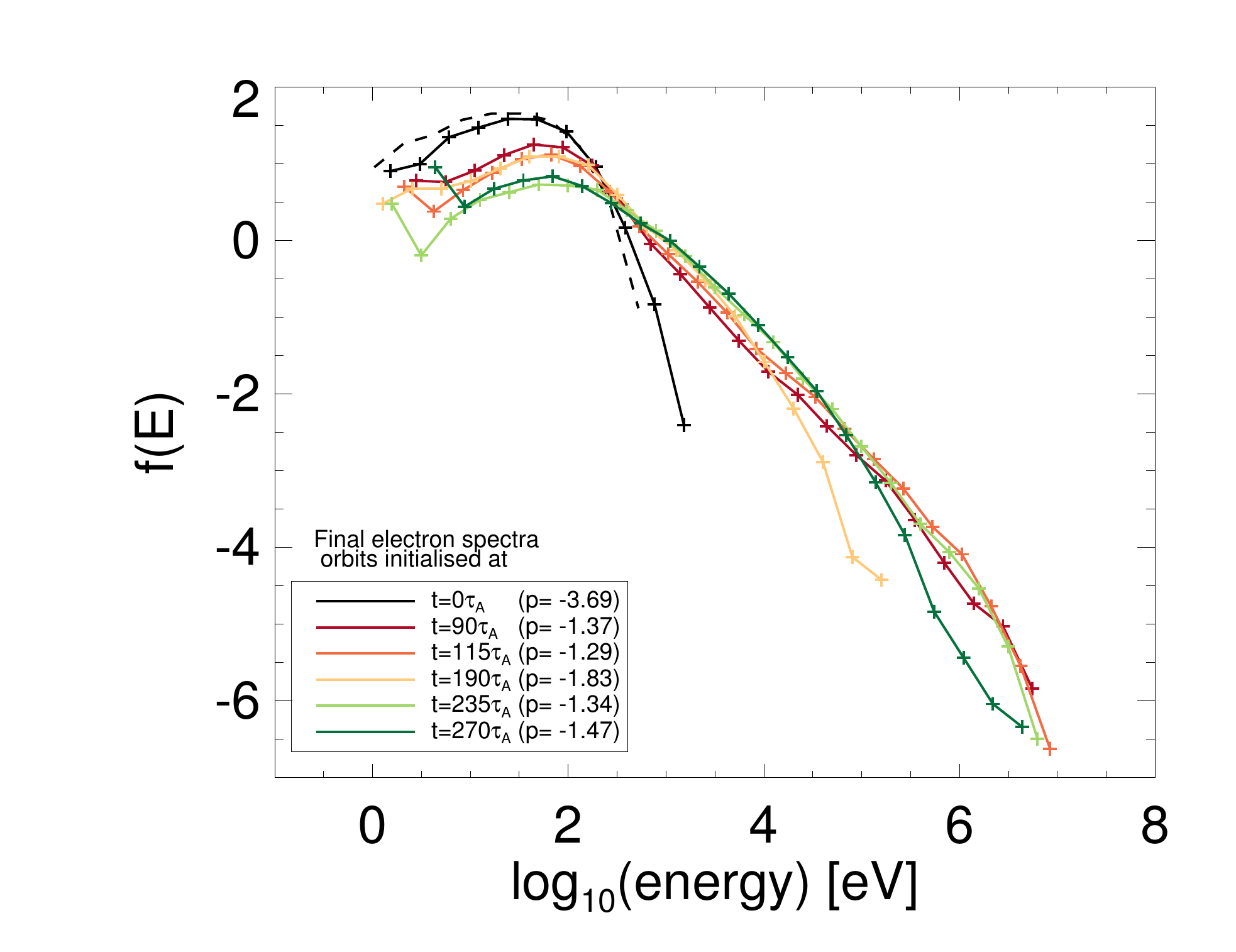}}}
  \subfloat[proton spectra]{\label{subfig:2LspecP}\resizebox{0.49\textwidth}{!}{\includegraphics[clip=true, trim=55 15 30 20]{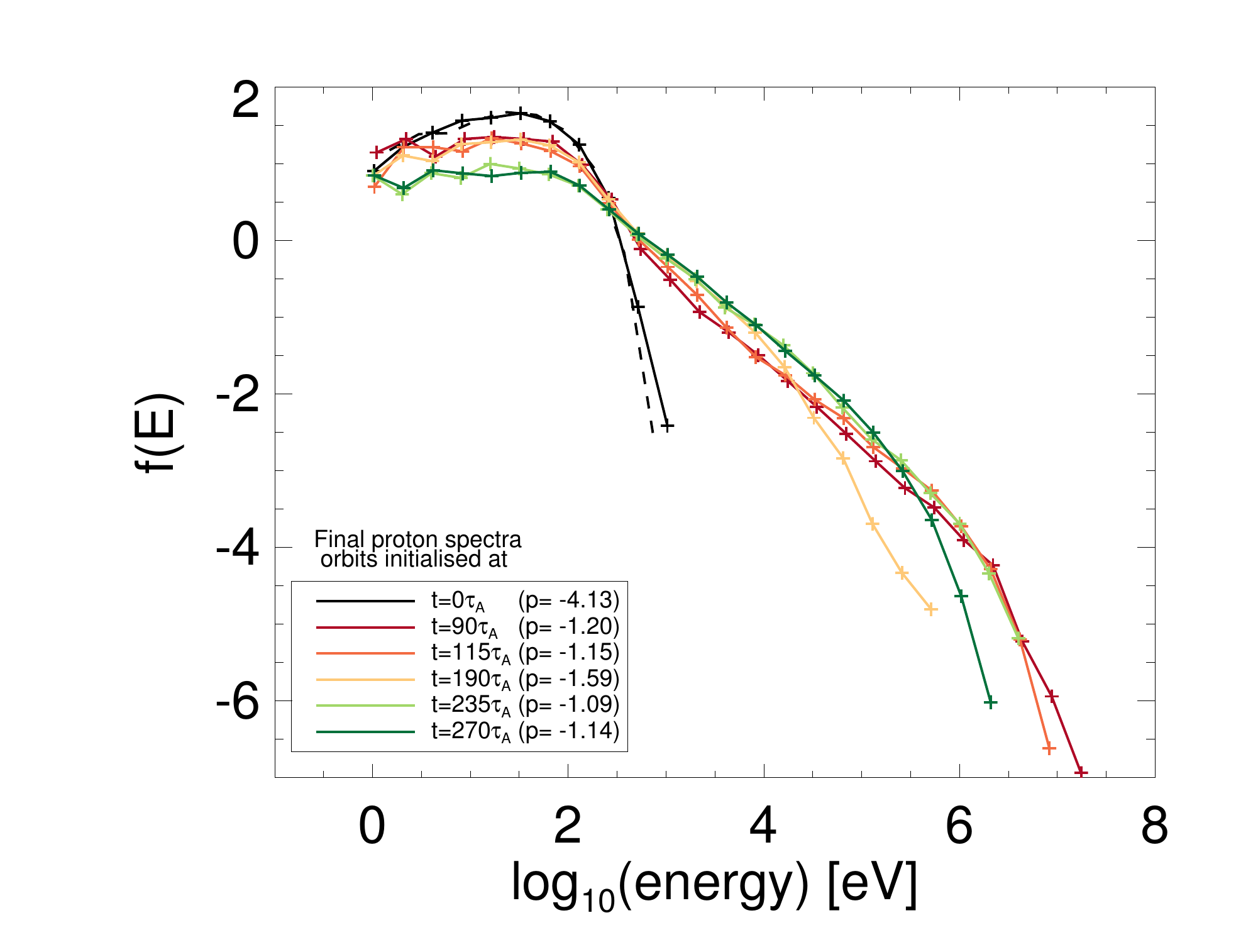}}}
  \caption{Interacting loops: temporal evolution of final energy spectra of \protect\subref{subfig:2Lspec} electrons and \protect\subref{subfig:2LspecP} protons, initiated at different snapshots during MHD experiment (where the fragmentation of one flux tube destabilises a second tube). \new{An initial energy spectrum is seen as a dashed histogram, while the coloured spectra represent the final energies for orbits started at different times. A power law has been fit to each spectra between $100\unit{eV}$ and $1\unit{MeV}$, with the power law index ($p$) and initiation time of each spectra given in the legend}.}
 \label{fig:twoloops_spec}
\end{figure*}

The final diagnostic information we will consider in this investigation are the energy distributions of the orbits in each of the six stages studied in Figs.~\ref{fig:twoloops_impact} and~\ref{fig:twoloops_impactP}. All spectra are presented together with an example of the initial Maxwellian energy spectrum for each population in Fig.~\ref{fig:twoloops_spec}.
 
\new{The electron spectra in Fig.~\ref{subfig:2Lspec} are generally softer than the proton spectra in Fig.~\ref{subfig:2LspecP}. In an attempt to quantify this, we fitted a simple power law to each distribution between energies of $100\unit{eV}$ and $1\unit{MeV}$, and recording the power law index ($p$) recovered with each spectrum. Apart from orbits using the initial snapshot (which are still close to a Maxwellian), all later spectra recover more negative values of $p$ for electrons than for protons, indicating sharper spectra and fewer non-thermal energy orbits. Power law index values lie between $-3.69$ and $-1.29$ for electrons and $-4.13$ and $-1.09$ for protons. The time-evolving spectra in Fig.~\ref{fig:twoloops_spec} are also strongly reminiscent of the energy partition of the MHD experiment seen in Fig.~\ref{fig:dEn2}.} That the initial stages closely match the original Maxwellian distribution is unsurprising; the same occurred for orbits in the single loop case which omitted background resistivity (dashed histograms in Fig.~\ref{fig:1Lt0En}). The onset of instability and fragmentation of the left-hand tube is accompanied by the appearance of a non-thermal tail in the energy distributions (at $t=90\tau_A$ and $t=115\tau_A$). At $t=190\tau_A$ (following the left-hand tube disruption but prior to that of the right-hand tube) fewer non-thermal orbits are generated, leaving a much more highly inclined slope with minimal activity above $100\unit{keV}$. A non-thermal tail returns in the final two cases following the disruption of the right-hand tube. Fewer MeV orbits are recorded in the final spectrum ($t=270\tau_A$) than that recovered at $t=235\tau_{A}$. This suggests that the system is once again reducing the amount of available current above the critical threshold with which to accelerate particles. At this stage, there is also reduced heating, as the system approaches a quasi-steady, lower energy state.

\section{Discussion}\label{sec:disc}
Our results highlight several key findings. The first of these is that the kink instability onset can accelerate significant numbers of particle orbits to high energies. This in itself is not a new result, and was studied in detail in the case of the kink instability onset in a single isolated flux tube in \citet{paper:GordovskyyBrowning2011,paper:GordovskyyBrowning2012}. What is new is our inclusion of a second flux tube, and how the particle acceleration characteristics change when moving from a single unstable flux tube to a cascade of flux tube disruptions. 

\subsection{Single thread destabilises}
Several facets of our findings reflect those of \citet{paper:GordovskyyBrowning2011,paper:GordovskyyBrowning2012}. Section~\ref{sec:oneunstable} \new{highlights a broadening of sites where accelerated orbits reach the domain boundaries} as the kink instability develops (and ultimately destabilises the flux tube). Resulting energy spectra also develop high-energy power-law-like tails which vary with time. We compare the final positions and energy gains using orbits which account for different profiles of magnetic resistivity (without changing the resistivity profile in the MHD simulation). The R2 resistivity profile considered in \citet{paper:GordovskyyBrowning2012} is comparable to the case where our orbits include both background and anomalous resistive effects, while the R1 profile is equivalent to the case where our orbits omit background resistive effects. The spectra recovered by the two resistivity profiles during Phases~2 \&~3 (seen in Figs.~\ref{subfig:1Lt1sp} \&~\ref{subfig:1Lt2sp}) bear similar hallmarks to the spectra recovered in Figs. 13 \& 14 of \citet{paper:GordovskyyBrowning2012}, where the inclusion of background resistivity typically enhances acceleration in the $1-100\unit{keV}$ energy range compared to the case with anomalous resistivity alone. We have shown that this difference is most pronounced during Phase~2 (Fig.~\ref{subfig:1Lt1sp}), but lessens during Phase~3 (Fig.~\ref{subfig:1Lt2sp}).

\new{The differences in the spectra in Phase~2 between cases with and without background resistivity are caused by the way the uniform and anomalous resistivities combine. Anomalous resistivity acts only on current above a critical threshold, switching off once current has dipped below $j_{\rm{crit}}$. Thus large regions of high but sub-critical current build up, to be acted upon by the background resistivity. These regions accelerate particles through the resulting parallel electric field, which are not as large as those created by anomalous resistivity, and hence yield more orbits at moderate energies.}

While the orbit response in Phases~2 \&~3 are broadly comparable to that seen in \citet{paper:GordovskyyBrowning2012}, Phase~1 provides additional insight in the single flux tube unstable case.  Prior to the kink instability onset, we completely disentangle the behaviour of the two resistivity profiles, during a Phase where current never exceeds the critical value and hence only the background resistivity is responsible for any acceleration. Asymmetric energised final positions at the upper and lower boundary (in Fig.~\ref{subfig:1Lt0pos} and~\ref{subfig:1Lt0posP}) are formed by the inclusion of background resistivity. Crucially, this not only occurs in the flux-tube in which the kink instability will occur, but also in a second neighbouring tube which remains stable throughout this experiment. Subsequent phases also display impact sites with these characteristics, but also modified due to anomalous resistive effects (not shown here). The halo and core structures result from the currents created in forming the flux tubes (e.g. Fig.~\ref{fig:jt0}) and exist throughout both tubes. Several recent papers have demonstrated similar findings in a variety of environments \citep{paper:Threlfalletal2015, paper:Threlfalletal2016b, paper:Threlfalletal2016a, paper:Threlfalletal2017a}, where even a weak but extended region of parallel electric field can be responsible for large energy gains. In our present normalisation, the loops are approximately $20$Mm in length. Even weak currents formed over this length, when subjected to weak background resistive effects, can yield significant particle energy gains. 

Another significant difference from \citet{paper:GordovskyyBrowning2012} is our choice of orbit boundary conditions. In this earlier work, three different boundary conditions were considered: fully transparent (as used here), partially reflective and fully reflective. We opted for the simplest of these: our objective was to study the impact of the time-variation of our simulations upon the orbits. Fully reflective boundary conditions would see the same set of particles respond to the different phases of our experiment, leaving it difficult to attribute specific features to the changing behaviour of the MHD background field. The partially reflective boundary conditions implemented in \citet{paper:GordovskyyBrowning2012}, mimic the effects of strong magnetic field convergence at the loop footpoints and ``effectively resulted in particles with the energies $>50-100\unit{keV}$ not being reaccelerated". The use of reflective boundary conditions for the orbits would perhaps explain the lack of any reported asymmetry in final boundary positions of high energy orbits (seen in e.g. Fig.~\ref{subfig:1Lt0pos} of our work) in the earlier paper. Our asymmetric energised final orbit positions are only present in cases where background resistivity alone is responsible for the acceleration. Many of the additional energised orbits resulting from background resistivity would not gain sufficient energy to enter the loss cone of the partially reflective boundary conditions used in \citet{paper:GordovskyyBrowning2012}, and hence would be reflected back into the simulation domain.

\subsection{Multi-thread cascade}
The primary goal of this investigation was to study how the acceleration picture changed when additional loop threads are destabilised by the eruption of a single thread. Our investigation considered a case containing two threads, where the MHD simulation utilised anomalous resistivity as the sole reconnection mechanism; no background resistivity was included (for clarity). The amount of particle acceleration, both in the number of highly accelerated orbits and the peak energy achieved by any orbit, appears closely tied to the heating rate (seen in Fig.~\ref{fig:dEn2}). This heating rate also reflects the reconnection rate, which is intrinsically linked to orbit energy gains \citep[e.g.][]{paper:Threlfalletal2016b}. The acceleration can be attributed to the availability of current above the critical value in these simulations. The final orbit positions associated with the instability of the left-hand thread remain closely associated with the footpoints of that initial tube, and slightly broaden over time (top row of Fig.~\ref{fig:twoloops_impact}). However, the disruption of the right-hand thread sees energised final positions `leak' out into regions formerly associated with the footpoints of the first thread, before spreading more widely along the upper and lower boundaries by the end of the experiment (e.g. bottom row of Fig.~\ref{fig:twoloops_impact}). This is consistent with a magnetic field which has become significantly braided over time. 

The recovered orbit spectra closely follow the heating rate seen in Fig.~\ref{fig:dEn2}. Figure~\ref{fig:twoloops_spec} shows that the high energy power law tail is hardest (i.e. most extended) when the kink instability develops the helical current sheet in the left-hand thread, before steepening and then returning to a similar trend during the eruption phase of the right-hand thread. Marginally higher energy orbits are associated with the earlier eruption spectrum, but this may simply result from our choice of number of orbits and initial conditions. 

That the spectra repeatedly harden then soften with each eruption is also noteworthy. Previous models of particle acceleration \citep[e.g.][]{paper:HannahFletcher2006,paper:GordovskyyBrowning2012,paper:Threlfalletal2015} often recover spectra which are directly tied to the reconnection electric field strength, steepening as the reconnection rate falls. Spectral softening and increased heating are often observed in the later stages of most solar flares \citep{review:Fletcheretal2011}. The link between spectra and reconnection electric field strength or rate also holds in this case, and results in the clear hard-soft-hard pattern we recover. However, a thread containing many loops \citep[e.g.][]{paper:Hoodetal2016} would yield a much less clear series of hard and soft spectra over time, depending on a number of factors, particularly how many and when each thread might destabilise.

We have also found that protons generally appear to produce slightly harder energy spectra than electrons; power law index values, fitted to spectra between energies of $100\unit{eV}$ and $1\unit{MeV}$, are more negative for electrons than protons except in the case where orbits are initiated at $t=0\tau_A$. The recovered power law indices lie well within the ranges described in \citet{paper:GordovskyyBrowning2012}. While relatively hard, similar index values have also been recovered in studies of a range of particle acceleration mechanisms \citep{paper:Baumannetal2013,paper:Stanieretal2012,paper:Threlfalletal2015}, and indeed agree with selected observational cases \citep{paper:Crosbyetal1993,paper:Kruckeretal2007,review:Hannahetal2011}. However, many more cases exist where such values are exceedingly low: in one example, \citet{paper:Hannahetal2008} analysed all available microflare spectra using RHESSI \citep{paper:Linetal2001}, commonly finding indices in the range of $4\,\--\,10$ with a median of $7$. As stated earlier, there are significant differences between our recovered spectra and observationally derived spectra that must be overcome before a detailed qualitative comparison may take place.

\new{Finally, we also note that rather than random initial orbit positions as used here, one could (in principle) produce a more `realistic' set of initial positions by weighting the number of orbits in specific regions according to the local MHD simulation plasma density. However, our MHD simulations begin with a uniform density profile, and hence this additional complexity in our approach would not make any difference to our results. Other factors are likely to also affect the resulting spectra, notably electromagnetic fields generated by the particles (which would predominantly affect the highest energy particles and act to soften the spectra), and collisions with background plasma (which should have little effect except near the loop footpoints where the density will be higher).}


\section{Conclusions and future work}\label{sec:conc}
We have studied particle orbit behaviour in a multi-threaded loop case for the very first time. 
An helical current sheet, surrounding a loop-thread along which the kink instability develops, is capable of accelerating significant quantities of electrons and ions (achieving upto approximately $10\unit{MeV}$ energies using present normalisation) towards the footpoints of the single thread. The fragmentation of this current sheet causes energetic particles to fill up the volume. Background resistive effects also generate significant particle acceleration (largely as a result of the currents formed by the structure of the loop threads themselves). However, the impact sites caused by background resistive effects differ at the thread footpoints, while anomalous resistive effects (acting above a critical current threshold) yield near-identical impact sites at both footpoints. In cases where a secondary thread eruption is triggered by the first eruption, a second acceleration event occurs, whose impact sites become mixed with the tangled remains of both threads, and whose spectra show a secondary hardening of high energy power law tails, with comparable particle energies achieved by the eruptions of either thread. Because of the field-line tangling and fragmented current structures forming in the second loop, the particles are accelerated throughout the volume of both loops

Several extensions of this work are readily apparent. This experiment shows that eruptions of multi-threaded loops can generate significant acceleration associated with each individual loop strand. How this presents in a case containing many threads (not all of which may disrupt), and resulting energetic particle impact sites and energy spectra would perhaps be more reflective of a true `nanoflare storm' model. Our choice of boundary conditions for each orbit do not necessarily reflect conditions in the solar atmosphere, particularly in regard of converging magnetic field at loop footpoints. \new{More work is also needed in order to compare energy distributions from test-particle models and observational spectra.}

\begin{acknowledgements}
The authors gratefully acknowledge the support of the U.K. Science and Technology Facilities Council. JT and AWH acknowledge the financial support of STFC through the Consolidated grant, ST/N000609/1, to the University of St Andrews. PKB acknowledges STFC support through ST/P000428/1 at the University of Manchester. This work used the DIRAC 1, UKMHD Consortium machine at the University of St Andrews and the DiRAC Data Centric system at Durham University, operated by the Institute for Computational Cosmology on behalf of the STFC DiRAC HPC Facility ({\url{www.dirac.ac.uk}}). This equipment was funded by a BIS National E-infrastructure capital grant ST/K00042X/1, STFC capital grant ST/K00087X/1, DiRAC Operations grant ST/K003267/1 and Durham University. DiRAC is part of the National E-Infrastructure. The MHD computations were carried out on the UKMHD consortium cluster housed at St Andrews and funded by STFC and SRIF.
\end{acknowledgements}

\bibliographystyle{aa}        
\bibliography{AA31915}          
\end{document}